%% file: Main_text_incl_figures.tex
\documentclass[aps,prb,reprint,amsmath,amssymb,superscriptaddress]{revtex4-1}
\usepackage{epsfig}
\usepackage{graphicx, subfigure}
\usepackage{color}

\usepackage{bm}
\usepackage{braket}
\usepackage{longtable}

\bibpunct{[}{]}{,}{n}{}{}

\usepackage{todonotes}
\usepackage{setspace}

\newcommand{\kb}{\mathbf{k}}

\newcommand{\eb}{\mathbf{E}}
\newcommand{\sigmab}{\boldsymbol{\sigma}}
\newcommand{\Lbhat}{\skew{-4}{\hat}{\mathbf{L}}}

\begin{document}

% Use the \preprint command to place your local institutional report
% number in the upper righthand corner of the title page in preprint mode.
% Multiple \preprint commands are allowed.
% Use the 'preprintnumbers' class option to override journal defaults
% to display numbers if necessary
%\preprint{}

%Title of paper
\title{Spin-orbit torques in locally and globally non-centrosymmetric crystals: Antiferromagnets and ferromagnets}

% repeat the \author .. \affiliation  etc. as needed
% \email, \thanks, \homepage, \altaffiliation all apply to the current
% author. Explanatory text should go in the []'s, actual e-mail
% address or url should go in the {}'s for \email and \homepage.
% Please use the appropriate macro foreach each type of information

% \affiliation command applies to all authors since the last
% \affiliation command. The \affiliation command should follow the
% other information
% \affiliation can be followed by \email, \homepage, \thanks as well.
\author{J. \v Zelezn\'y}
\affiliation{Institute of Physics ASCR, v.v.i., Cukrovarnick\'a 10, 162 53
Praha 6, Czech Republic}
\affiliation{Faculty of Mathematics and Physics, Charles University in Prague,
Ke Karlovu 3, 121 16 Prague 2, Czech Republic}
\author{H. Gao}
\affiliation{Institut f\"ur Physik, Johannes Gutenberg Universit\"at Mainz, 55128 Mainz, Germany}
\affiliation{Department of Physics, Texas A\&M University, College Station, Texas 77843-4242, USA}
\author{Aur\'elien Manchon}
\affiliation{Physical Science and Engineering Division, King Abdullah University of Science and Technology (KAUST), Thuwal 23955-6900, Kingdom of Saudi Arabia}
\author{Frank Freimuth}
\affiliation{Peter Gr\"unberg Institut and Institute for Advanced Simulation, Forschungszentrum J\"{u}lich and JARA, 52425 J\"{u}lich, Germany}
\author{Yuriy Mokrousov}
\affiliation{Peter Gr\"unberg Institut and Institute for Advanced Simulation, Forschungszentrum J\"{u}lich and JARA, 52425 J\"{u}lich, Germany}
\author{J. Zemen}
\affiliation{Department of Physics, Blackett Laboratory, Imperial College London, London SW7 2AZ, United Kingdom}
\author{J. Ma\v{s}ek}
\affiliation{Institute of Physics ASCR, v.v.i., Na Slovance 1999/2,
182 21 Praha 8, Czech Republic}
\author{Jairo Sinova}
\affiliation{Institut f\"ur Physik, Johannes Gutenberg Universit\"at Mainz, 55128 Mainz, Germany}
\affiliation{Institute of Physics ASCR, v.v.i., Cukrovarnick\'a 10, 162 53
Praha 6, Czech Republic}
\author{T. Jungwirth}
\affiliation{Institute of Physics ASCR, v.v.i., Cukrovarnick\'a 10, 162 53
Praha 6, Czech Republic} 
\affiliation{School of Physics and
Astronomy, University of Nottingham, Nottingham NG7 2RD, United Kingdom}

%\email[]{Your e-mail address}
%\homepage[]{Your web page}
%\thanks{}
%\altaffiliation{}

%Collaboration name if desired (requires use of superscriptaddress
%option in \documentclass). \noaffiliation is required (may also be
%used with the \author command).
%\collaboration can be followed by \email, \homepage, \thanks as well.
%\collaboration{}
%\noaffiliation

\date{\today}

\begin{abstract}
One of the main obstacles that prevents practical applications of antiferromagnets is the difficulty of manipulating the magnetic order parameter. Recently, following the theoretical prediction [J. \v Zelezn\'y et al., PRL 113, 157201 (2014)], the electrical switching of magnetic moments in an antiferromagnet has been demonstrated [P. Wadley et al., Science 351, 587 (2016)]. The switching is due to the so-called spin-orbit torque, which has been extensively studied in ferromagnets. In this phenomena a non-equilibrium spin-polarization exchange coupled to the ordered local moments is induced by current, hence exerting a torque on the order parameter. Here we give a general systematic analysis of the symmetry of the spin-orbit torque in locally and globally non-centrosymmetric crystals. We study when the symmetry allows for a nonzero torque, when is the torque effective, and its dependence on the applied current direction and orientation of magnetic moments. For comparison, we consider both antiferromagnetic and ferromagnetic orders. In two representative model crystals we perform microscopic calculations of the spin-orbit torque to illustrate its symmetry properties and to highlight conditions under which the spin-orbit torque can be efficient for manipulating  antiferromagnetic moments. 
\end{abstract}

% insert suggested PACS numbers in braces on next line
% insert suggested keywords - APS authors don't need to do this
%\keywords{}

%\maketitle must follow title, authors, abstract, \pacs, and \keywords
\maketitle

% body of paper here - Use proper section commands
% References should be done using the \cite, \ref, and \label commands
\section{Introduction}
% Put \label in argument of \section for cross-referencing
%\section{\label{}}

Antiferromagnets (AFMs) have so far found little applications as active components of devices primarily due to their lack of net magnetization. With the development of spintronics, however, the net magnetization that couples strongly to the magnetic field becomes less important. In the latest generation of magnetic random access memories (MRAMs), for example, magnetic fields are used neither for writing nor for reading. Since AFMs possess a long range magnetic order just like ferromagnets (FMs), they have been recently explored as new materials for spintronics (see Refs. \cite{Gomonay2014,Macdonald2011,Jungwirth2015} for recent reviews of antiferromagnetic spintronics). In particular, they could in principle be used for solid state memories in which bits of information are represented by the direction of the magnetic order parameter, similarly to FMs. Such memory functionalities were experimentally demonstrated in AFM tunneling \cite{Park2011} and ohmic devices \cite{Xavi2014,Moriyama2015,Wadley2016}.

Compared to FMs, AFMs have several potential advantages. They are insensitive to large magnetic fields and do not produce any stray fields. This makes them more challenging from an experimental and technological point of view, but it can also be an advantage. Stray fields can cause problems in densely packed devices, and the sensitivity to external magnetic fields means that a FM memory can be accidentally rewritten by external magnetic fields. AFM memory, on the other hand, is much less sensitive to external magnetic fields. For example, a memory based on FeRh could not be erased by fields as high as 9 T \cite{Xavi2014}. Another advantage is that dynamics of magnetic moments in AFMs is much faster than in FMs. Switching of the AFM order parameter on a ps timescale was demonstrated, e.g., in a laser-induced heating experiment \cite{Kimel2004}.

A remarkable feature of AFMs is also the wide range of available AFM materials. This holds especially for semiconductors. FM semiconductors have attracted a lot of interest in the past since they enable the combination of spintronic and microelectronic functionalities. Yet, despite intensive research, FM semiconductors remain rare and tend to have Curie temperatures too low for practical applications. AFM semiconductors on the other hand are more common and tend to have magnetic order persisting above room temperature \cite{Fina2014,Kriegner2015,LiMnAs,Jungwirth2015}. Materials that combine antiferromagnetism with ferroelectricity \cite{Sando2013} or the parent compounds of the high $T_c$ superconductors \cite{superconductivityscience} further highlight the broad and diverse range of AFMs.

For microelectronic memory and logic applications of AFMs, two basic functionalities have to be available: a method for detecting and manipulating electrically the magnetic order parameter. For readout, the anisotropic magnitoresistance (AMR) effect \cite{Fina2014,Wang2014,Kriegner2015} and its tunneling counterpart TAMR have been demonstrated \cite{Park2011}. While AMR is usually rather small, with typical magnetoresistance ratios around a few percent, a $\sim 100\%$ TAMR has been already achieved, albeit at low temperatures.

Manipulating the magnetic order parameter in AFMs by practical means has been a major challenge. AFMs can be controlled by external magnetic fields, but this is impractical since it typically requires very large fields. The lowest uniform static field that can reorient an AFM is the so-called spin-flop field, which is proportional to $\sqrt{H_{J}H_{\text{an}}}$, where $H_J$ is the inter-sublattice exchange field and $H_{\text{an}}$ is the anisotropy field. Since the exchange interaction is typically much larger than the anisotropy, the spin-flop fields are large compared to FMs. Instead, an auxiliary exchanged-coupled FM layer is often used \cite{Scholl2004,Park2011,Fina2014}, which makes manipulation possible by smaller fields. This only works for thin AFM layers though, and it is highly dependent on interface properties. 

While FMs can be manipulated by external magnetic fields, in microelectronic devices a direct electrical manipulation offers a more scalable approach. This is usually achieved using the so-called spin-transfer torque \cite{Slonczewski1989,Ralph2008}. This torque occurs due to the absorption of angular momentum from a spin-polarized current generated by a fixed FM polarizer. On the other hand, due to spin-orbit coupling, a torque can be generated without the injection of a spin-current from the FM polarizer \cite{Bernevig2005,Chernyshov2009,Manchon2008,Garate2009,Abiague2009,Gambardella2011,Kurebayashi2014,Fang2011,Miron2011,Liu2012,Li15}. Such torque is usually called a spin-orbit torque. In FMs it requires a broken inversion symmetry and can therefore occur either in crystals with no inversion symmetry in the unit cell or in heterostructures, where inversion symmetry is broken structurally.

Because of the insensitivity of AFMs to external fields, the electrical manipulation of AFMs is even more desirable. To manipulate a collinear AFM effectively, a staggered magnetic field (i.e., a field that is opposite on the two sublattices) is needed. In  Ref. \cite{Gomonay2012} it was shown that the analogue of the FM spin-transfer torque in AFMs can generate effective fields that are staggered. However this requires very thin layers \cite{Reichlova2015}.

Ref. \cite{Zelezny2014} proposed that in bulk AFMs with specific symmetries, electrical current can create a torque by a similar mechanism  to the spin-orbit torque in FMs. The work also showed that the effective field generating the torque can be staggered and the corresponding non-staggered torque can thus be effective for manipulating AFMs. Switching of an AFM based on  predictions in Ref. \cite{Zelezny2014} was recently experimentally observed in AFM CuMnAs \cite{Wadley2016}. This opens up a way to applications of AFMs. The current densities needed for switching in Ref. \cite{Wadley2016} were comparable to current densities in FM spin-torque MRAMs.

In this manuscript we theoretically study the nature and characteristics of spin-orbit torques in AFMs in a systematic way. We give a general symmetry analysis for locally and globally non-centrosymmetric crystals and considering both the AFM and FM order. We determine when the torque can exist, when it is effective for manipulating the magnetic order in AFMs, and also what form the torque has. In Ref. \cite{Zelezny2014} the spin-orbit torque was calculated for two representative tight-binding models, one describing a three-dimensional (3D) lattice of $\text{Mn}_2\text{Au}$ and the other one representing a two-dimensional (2D) crystal with Rashba spin-orbit coupling. $\text{Mn}_2\text{Au}$ served as a model AFM system with globally centrosymmetric and locally non-centrosymmetric crystal structure and inversion-partner lattice sites occupied by the two spin-sublattices. In this model system, the so called field-like torque, driven by a staggered, magnetization-independent current-induced field, is effective. (Note that the relevant crystal symmetries of $\text{Mn}_2\text{Au}$ are the same as  those of the recently experimentally studied CuMnAs  and that the calculated magnitudes of the spin-orbit torques in $\text{Mn}_2\text{Au}$ and CuMnAs are also comparable [7].) On the other hand, the so called (anti)damping-like torque, driven by a staggered, magnetization-dependent effective field, was found to be the effective torque component in the 2D AFM crystal with a global inversion asymmetry modeled by the Rashba Hamiltonian. 

Here we calculate all spin-orbit torque components in both models which allows us to generalize the result of Ref. \cite{Zelezny2014}: All torque components driven by fields that are an even function of the sublattice magnetization are effective in the AFM 3D $\text{Mn}_2\text{Au}$ model while torques driven by fields that are odd in magnetization are effective in the AFM 2D Rashba model. The calculations also reveal that the angular dependences of the current-induced fields with respect to the applied current direction and the direction of magnetic moments are similar in the two model systems, due to similarities in the relevant symmetries of the two model crystals. Numerical and analytical calculations of the spin-orbit torque in the two tight-binding models are complemented by ab initio density-functional-theory (DFT) calculations and results for the AFM order are compared to calculations assuming the FM order in the same model crystals.

Our paper is organized as follows: In Sec. \ref{sec:models} we describe the two tight-binding models and the linear response formalism used for calculating the spin-orbit torque. In Sec. \ref{sec:symmetry} we discuss the symmetry of the spin-orbit torque and apply the general symmetry arguments to our two models. A detailed derivation of symmetry properties of the spin-orbit torque is given in Appendix \ref{appendix:symmetry}. In Sec. \ref{sec:results} we show the results of analytical and numerical calculations of the spin-orbit torque in the two models. In Sec. \ref{sec:discussion} we discuss the results, in particular summarize the symmetry considerations. 

\section{Models}\label{sec:models}
In some materials electrical current can induce non-equilibrium spin-polarization due to spin-orbit coupling \cite{Edelstein1990,Kato2004,Silov2004,Ganichev2004,Wunderlich2005}. This effect is called the inverse spin-galvanic effect or the Edelstein effect. For the presence of non-vanishing net spin-polarization (i.e., integrated over the whole unit cell) a broken inversion symmetry is needed. In FMs, due to exchange interaction between carrier spins and magnetic moments, the current-induced spin-polarization (CISP) will exert a torque on the magnetization. This effect is the spin-orbit torque. In AFMs the effect is similar. Since the carrier--magnetic moment exchange interaction is short-range, spin-polarization generated by the electrical current on a sublattice will interact primarily with the magnetic moments on that sublattice. To evaluate the spin-orbit torque in AFMs we thus have to calculate the CISP locally on each magnetic sublattice. 

Note that in the spintronics community two different effects are termed as the spin-orbit torque. Apart from the effect discussed here, there exists also a torque generated in heavy metal/FM heterostructures due to spin Hall effect. Lateral electrical current generates spin current in the perpendicular direction due to the spin Hall effect, which flows in the FM and exerts a torque via the spin-transfer torque mechanism. Since the heterostructures have broken inversion symmetry, the torque due to inverse spin galvanic effect coexists with the spin Hall torque, rendering the entire physics quite complex to analyze (see for instance Ref. \cite{freimuth2014}, where both mechanisms are included). We only consider bulk systems in which the spin Hall effect does not generate any torque.

To calculate the CISP $\delta \mathbf{S}_a$ ($a$ denotes the sublattice) we use the Kubo linear response formalism. We can define a response tensor $\chi_{a}$ such that $\delta \textbf{S}_{a} = \chi_{a} \mathbf{E}$, where $\mathbf{E}$ is the electrical field. Assuming constant quasiparticle broadening $\Gamma$ and weak disorder (i.e., small $\Gamma$), the tensor $\chi_a$ can be expressed as a sum of three terms \cite{Li15},

\begin{gather}
\chi_a  = \chi_{a}^\text{I} + \chi_{a}^\text{II(a)} + \chi_{a}^\text{II(b)},\label{eq:Kubo}\\    
  \chi_{a,ij}^\text{I} = -\frac{e\hbar}{2\Gamma}
  \sum_{\kb,n} \Bra{\psi_{n\mathbf{k}}}\hat{S}_{a,i}\Ket{\psi_{n\mathbf{k}}}\Bra{\psi_{n\mathbf{k}}}\hat{v}_j\Ket{\psi_{n\mathbf{k}}}\notag\\
  \times\delta(\varepsilon_{\mathbf{k}n}-E_F),\label{Boltzmann}\\
  \chi_{a,ij}^\text{II(a)} = {e\hbar} \sum_{\mathbf{k},n\neq m} \text{Im} [\Bra{\psi_{n\mathbf{k}}}\hat{S}_{a,i}\Ket{\psi_{m\mathbf{k}}}\Bra{\psi_{m\mathbf{k}}}\hat{v}_j\Ket{\psi_{n\mathbf{k}}}]\notag\\
  \times \frac{\Gamma^2-(\varepsilon_{\kb n}-\varepsilon_{\kb m})^2}{[(\varepsilon_{\kb n}-\varepsilon_{\kb m})^2+\Gamma^2]^2}(f_{\kb n}-f_{\kb m}),\label{eq:Kubo1}\\
  \chi_{a,ij}^\text{II(b)} = {2e\hbar} \sum_{\mathbf{k},n\neq m} \text{Re} [\Bra{\psi_{n\mathbf{k}}}\hat{S}_{a,i}\Ket{\psi_{m\mathbf{k}}}\Bra{\psi_{m\mathbf{k}}}\hat{v}_j\Ket{\psi_{n\mathbf{k}}}] \notag\\
  \times \frac{\Gamma(\varepsilon_{\kb n}-\varepsilon_{\kb m})}{[(\varepsilon_{\kb n}-\varepsilon_{\kb m})^2+\Gamma^2]^2}(f_{\kb n}-f_{\kb m}),\label{eq:Kubo2}
\end{gather}
where $n,m$ are band indices, $\psi_{n \kb}$ and $\varepsilon_{n \kb}$ denote Bloch eigenfunctions and eigenvectors respectively, $E_{F}$ is the Fermi energy, $f_{\kb,n}$ the Fermi-Dirac distribution function, $\hat{\mathbf{v}}$ is the velocity operator, $e$ is the (positive) elementary charge and $\hat{\mathbf{S}}_a$ is the spin-operator projected on sublattice $a$. Throughout this text we use a dimensionless spin-operator, i.e., for one electron $\hat{\mathbf{S}} = \sigmab$, where $\sigmab$ is a vector of Pauli matrices. The $\kb$ sums  run over the first Brillouin zone. $\chi_a^\text{I}$ is called the intraband term and $\chi_{a}^\text{II(a)}$, $\chi_{a}^\text{II(b)}$ are the interband terms. In the limit of $\Gamma \xrightarrow{} 0$, $\chi_a^\text{I}$ diverges, $\chi_{a}^\text{II(a)}$ is a constant and $\chi_{a}^\text{II(b)}$ is zero. These equations are the same as in Ref. \cite{Li15}, except we replace the spin-operator by the spin-operator projected on a sublattice. We calculate the CISP for the AFM spin-sublattices. However, the same formalism applies also for any sublattice in a FM or a non-magnetic material.

\begin{figure}[h]
  \includegraphics[width=8.6cm]{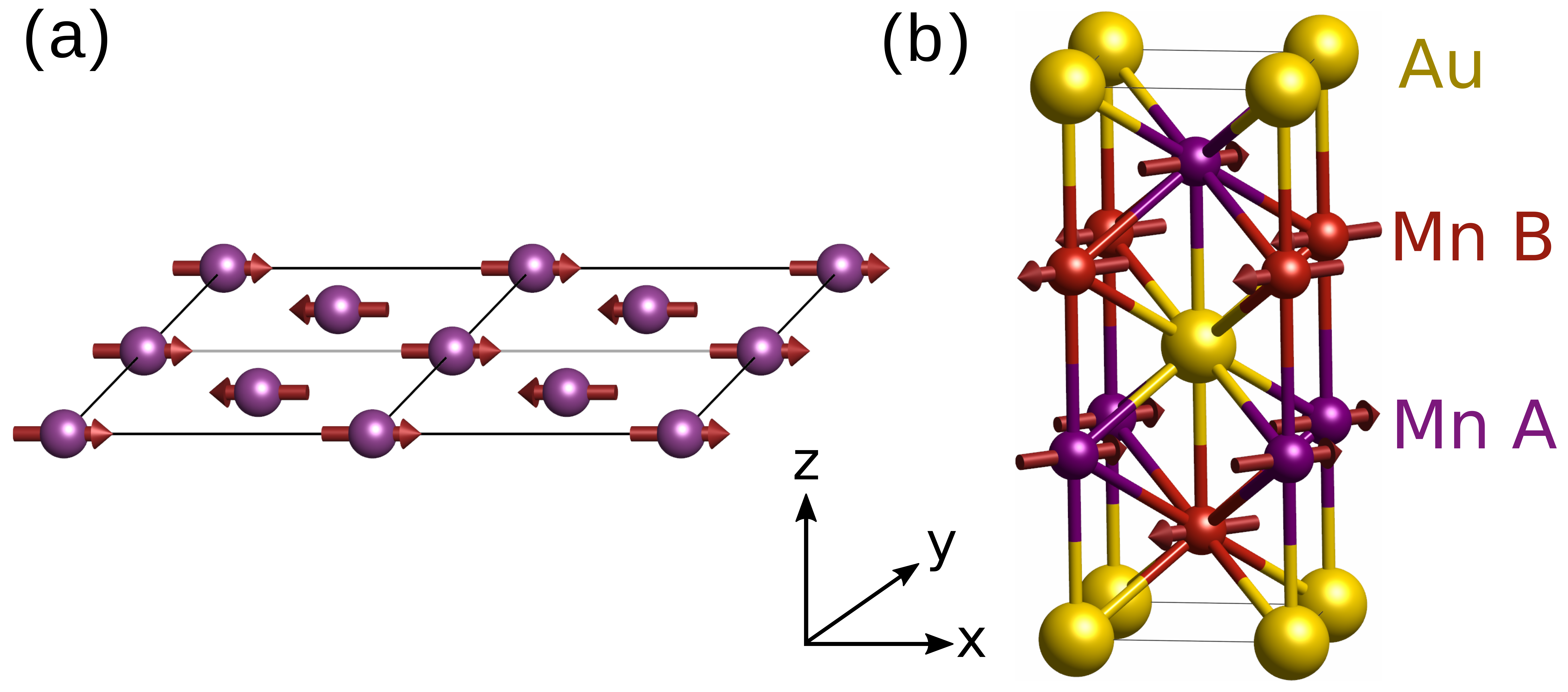}
  \caption{(Color online) Crystal structure of two model AFMs. (a) Crystal structure of the AFM 2D Rashba model. (b) Crystal structure of AFM $\text{Mn}_2\text{Au}$. Note that the unit cell shown is the conventional unit cell, which is as large as the primitive unit cell. All of the atoms with the same color are connected  by a translation and are thus equivalent.}
  \label{fig:structures}
\end{figure}

We calculated the CISP for the two tight-binding models from Ref. \cite{Zelezny2014}. For completeness we give here a description of the models. The first one is a 2D tight-binding model with Rashba spin-orbit coupling, which simulates the structural inversion asymmetry at a surface or an interface. The model was chosen as a simplest AFM model in which the spin-orbit torque is expected. We consider a square AFM lattice (see Fig. \ref{fig:structures}(a)), where the $d$-orbital local magnetic moments are treated classically and only the conduction \emph{s} electrons are treated quantum mechanically. The Hamiltonian can be written as

\begin{align}
 H = \sum_{<ij>}J_{dd} \hat{\mathbf{M_i}}\cdot\hat{\mathbf{M_j}} + H^{tb} + \sum_i J_{\text{sd}} \hat{\mathbf{S}}_i\cdot\hat{\mathbf{M}}_i + H_R.
 \label{eq:2Dmodel}
\end{align}

Here the indices $i,j$ correspond to lattice sites; $\hat{\mathbf{M}}_i,\hat{\mathbf{M}}_j$ are directions of magnetic moments, $J_{dd}$ and $J_{\text{sd}}$ are the exchange constants for exchange interaction between the magnetic moments, and between the magnetic moments and conduction electron spins, respectively. $H^{tb}$ contains the nearest neighbor hoppings. $H_R$ is the Rashba spin orbit coupling, given by
\begin{align}
H_R = \frac{\alpha}{2a_l}&\sum_j [(c_{j\uparrow}^{\dagger} c_{j+\delta_x\downarrow} - c_{j\downarrow}^{\dagger} c_{j+\delta_x\uparrow}) -\notag\\
&- i(c_{j\uparrow}^{\dagger} c_{j+\delta_y\downarrow} + c_{j\downarrow}^{\dagger} c_{j+\delta_y\uparrow}) + \mbox{H.c.}],
\end{align}
where $\alpha$ is the Rashba parameter, $a_l$ the lattice constant, $c_j^\dagger$, $c_j$ are the creation and annihilation operators for electron on site $j$, and $j+\delta_x$, $j+\delta_y$ are nearest neighbors along the $x$ and $y$ directions, respectively. Ref. \cite{Zelezny2014} shows the band structure of this Hamiltonian. In all calculations we set $t=3\ \text{eV}$, $J_{\text{sd}}=1\ \text{eV}$, and  $\alpha=0.1\ \text{eV}$, where $t$ is the hopping parameter. Unless stated otherwise, the Fermi level is set to $E_F=-2\ \text{eV}$.

The torque is given by
\begin{align}
 \mathbf{T}_a = \mathbf{M}_a \times \mathbf{B}_a,\label{eq:torque}
\end{align}
where $\mathbf{M}_a$ is the magnetic moment on sublattice $a$ and $\mathbf{B}_a$ is the effective current-induced field, which for this model is given by \cite{Gambardella2011}
\begin{align}
 \mathbf{B}_a =-J_{\text{sd}} \frac{\delta \mathbf{S}_a}{M_a},
 \label{eq:eff_field}
\end{align}
where $M_a$ is the magnitude of magnetic moment on sublattice $a$.

The second model describes a 3D AFM $\text{Mn}_2\text{Au}$. The crystal structure of $\text{Mn}_2\text{Au}$ is shown in Fig. \ref{fig:structures}(b). It is a collinear high N\'eel temperature AFM that has recently been identified as a promising material for AFM spintronics \cite{Shick2010,Barthem2013,Klaui2015}. We describe $\text{Mn}_2\text{Au}$ by an empirical tight-binding Slater-Koster model with \emph{s,p} and \emph{d} electrons for each atom. We use the tight-binding parameters for single-element metals from Ref. \cite{Shi2004} as a starting point and improve them so that the model agrees with the DFT calculation \cite{Zelezny2014}. (See Ref. \cite{Zemen2014} for details of the method and the procedure for obtaining the tight-binding parameters.) The DFT calculation was done using the full-potential all-electron code Wien2k \cite{wien2k}. To improve the description of the Mn \emph{d} states, we used the LDA+U method with U=4.63 eV and J=0.54 eV \cite{Wadley2013}.
 
For the tight-binding calculations of the CISP we add a $\mathbf{k}$-independent on-site spin-orbit coupling for both Mn and Au atoms with parameters obtained from atomistic Hartree-Fock calculations. The tight-binding model is not expected to be quantitatively as accurate as DFT calculation, however, it can be used to illustrate the origin and the symmetries of the spin-orbit torque in the AFM $\text{Mn}_2\text{Au}$ crystal. A quantitative comparison to DFT spin-orbit torque calculations is presented in Sec. \ref{sec:discussion}.
 
In the DFT calculation it is possible to evaluate the effective field or directly the torque using the space-dependent exchange field \cite{Garate2009,freimuth2014}. In the case of the tight-binding calculation, we obtain only the CISP. To get an estimate of the effective field we can still use Eq. \eqref{eq:eff_field}, which corresponds to taking a spatial average of the exchange field. In Ref. \cite{Zelezny2014} the carrier--magnetic moment exchange constant was set to $J_\text{sd}=1\ \text{eV}$, which is a typical value estimated for transition metals \cite{Haney2014}.
 
In Ref. \cite{Zelezny2014}, only the terms $\chi_a^{II(a)}$ for the 2D Rashba model and $\chi_a^{I}$ for the $\text{Mn}_2\text{Au}$ model were considered, respectively. Here we take into account all three terms for both models. Since we are primarily interested in the small $\Gamma$ limit, we mostly focus on terms $\chi_a^{I}$, $\chi_a^{II(a)}$, but the term $\chi_a^{II(b)}$ is also discussed. 
% \begin{figure}
%   \includegraphics[scale=0.3]{./gridp.png}
%   \caption{\label{fig:2dmodel}}
% \end{figure}
% 
% \begin{figure}
%   \includegraphics[scale=0.25]{./$\text{Mn}_2\text{Au}$2.png}
%   \caption{\label{fig:$\text{Mn}_2\text{Au}$}Crystal structure of $\text{Mn}_2\text{Au}$.}
% \end{figure}

\section{Symmetry considerations}
\label{sec:symmetry}
Symmetry is crucial for understanding when the spin-orbit torque can exist and what form it has. In Appendix \ref{appendix:symmetry} we give a derivation of the symmetry properties of the tensor $\chi_a$. Here we summarize the main results and apply them to our two models. The following analysis applies both to the effective field and the CISP because they have the same symmetry properties. Since spin-orbit torque is a non-equilibrium process that includes dissipation, the tensor $\chi_a$ does not have a simple behavior under time-reversal. To deal with this problem we separate the tensor into a part even in magnetic moments and a part odd in magnetic moments 
\begin{align}
 \chi_a^{\text{even}}(\big[\mathbf{M}]) = \big[\chi_a([\mathbf{M}]) + \chi_a([-\mathbf{M}])\big]/2,\\
 \chi_a^{\text{odd}}(\big[\mathbf{M}]) = \big[\chi_a([\mathbf{M}]) - \chi_a([-\mathbf{M}])\big]/2,
\end{align}
where $[\mathbf{M}] = [\mathbf{M}_A,\mathbf{M}_B,\dots]$ denotes the directions of all magnetic moments in the magnetic unit cell. As shown in the Appendix \ref{appendix:symmetry} it holds that 
\begin{align}
 \chi_a^{\text{even}} &= \chi_a^{I}+\chi_{a}^{II(b)},\\
 \chi_a^{\text{odd}} &= \chi_{a}^{II(a)}.
\end{align}
In Appendix \ref{appendix:symmetry}, the following rules are derived for the transformation of $\chi_a$ under symmetry operation $R$
\begin{align}
 \chi^{\text{even}}_{a'} &= \det(D) D \chi^{\text{even}}_a D^{-1},\label{eq:trans_f1}\\
 \chi^{\text{odd}}_{a'} &= \pm \det(D) D \chi^{\text{odd}}_a D^{-1}\label{eq:trans_f2},
\end{align}
where $a'$ is the sublattice to which the sublattice $a$ transforms under symmetry operation $R$ and $D$ is a matrix representing the symmetry operation in real space as defined by Eq. \eqref{eq:matrix_D}. The plus sign in Eq. \eqref{eq:trans_f2} corresponds to a symmetry operation that does not contain time-reversal and the minus to a symmetry operation that contains time-reversal. These rules apply for any form of magnetic order as well as for nonmagnetic crystals. The same rules also apply for the tensor $\chi$, which describes the net CISP.

Basic symmetry rules can be inferred from Eqs. \eqref{eq:trans_f1} and \eqref{eq:trans_f2}. If the system has an inversion symmetry
\begin{align}
 \chi_{a'} = -\chi_{a}.
\end{align}
If also inversion transforms the sublattice $a$ into itself, then  there can be no CISP on the sublattice  $a$. We therefore reach an important conclusion: for the existence of the CISP (and thus also the spin-orbit torque) on sublattice $a$, the inversion symmetry has to be \emph{locally} broken, i.e., there must be no inversion center in the sublattice $a$. This means that current can generate spin-polarization even in a material that has global inversion symmetry if inversion symmetry is broken locally. However, it is also important to note that if the inversion symmetry is locally broken, the CISP can still vanish due to other symmetries. For example, a diamond lattice has a global inversion symmetry, but the two different lattice sites in the diamond unit cell have inversion symmetry locally broken. Without any strain, the CISP will nevertheless vanish. However, when a uniaxial strain is present in the diamond lattice a CISP with opposite sign on the two different sites will appear \cite{ciccarelli2015}.

In the 2D Rashba model, the inversion symmetry is broken globally due to the structural asymmetry of the assumed layered system. In the AFM $\text{Mn}_2\text{Au}$ crystal, the inversion symmetry is broken by the magnetic order since the inversion partner lattice sites are occupied by Mn atoms with opposite moments. Even without magnetic moments, however, the inversion symmetry is locally broken for each sublattice. This can be seen in Fig. \ref{fig:structures}(b) and is discussed in more detail in Sec. \ref{sec:discussion} (see also Fig. \ref{fig:Mn2Au_inv}).

Of particular interest in the case of AFMs is to determine how the CISPs on the AFM spin-sublattices are related. This is because for the spin-orbit torque to be efficient, the current induced effective magnetic field and thus also the CISP have to be staggered. Since the exchange interaction is much larger than any field typically acting on AFMs, we assume that during any dynamics the two magnetic moments stay approximately collinear (although any dynamics of the AFM order parameter induces a small magnetization). Then in the AFM 2D Rashba model, a simultaneous translation and time-inversion will always be a symmetry of the model that transforms one AFM spin-sublattice into the other. For such a symmetry operation, $D=I$, where $I$ is the identity matrix, and therefore
\begin{align}
 \chi_{A}^{\text{even}} &= \chi_{B}^{\text{even}},\\
 \chi_{A}^{\text{odd}} &= -\chi_{B}^{\text{odd}}.
\end{align}
This implies that the efficient torque driven by a staggered field is generated by the odd component of the response tensor.

In the $\text{Mn}_2\text{Au}$ type of crystal, the AFM spin-sublattices are not connected by translation. Instead they are connected by inversion around the unit cell center so that a combination of inversion and time-reversal is a symmetry of the model. Since in this case $D=-I$, we find
\begin{align}
 \chi_{A}^{\text{even}} &= -\chi_{B}^{\text{even}},\\
 \chi_{A}^{\text{odd}} &= \chi_{B}^{\text{odd}}
\end{align}
and now it is the even component of the response tensor that generates the staggered CISP.
 The two models illustrate a general phenomenology of CISPs in collinear AFMs, in which the two AFM spin-sublattices are typically connected either by a translation or by an inversion.

By considering the magnetic space group of a given material, one can find using the Eqs. \eqref{eq:trans_f1} and \eqref{eq:trans_f2} the most general form of the tensor $\chi_a$ as well as relations between tensors $\chi_a$ on different sublattices. Note that for the CISP projected on a sublattice it is not enough to consider the point group of the crystal because then the information on the relationship between the sublattices would be lost.  We provide a free program which outputs the symmetry of the CISP for any type of crystal and magnetic structure \cite{symcode}. See the Appendix \ref{appendix:code} for a brief description of the code. Symmetry of the tensors, which describe the global spin-orbit torque can be found in Ref. \cite{Wimmer2016} for every magnetic point group. These also apply for the local spin-orbit torque, if the local magnetic point group is used.

To describe the dependence of the CISP on the direction of magnetic moments, it is useful to expand the linear response tensor in powers of magnetic moments. In general $\chi_a$ depends on the directions of all magnetic moments in the system. We consider only FMs and collinear two-sublattice AFMs. We again assume that the magnetic moments will always stay approximately collinear. Since the intra-spin-sublattice exchange is typically very large, we also assume that the magnitude of the spin-sublattice magnetic moments will not change during dynamics. Then $\chi_a$ will be a function of only the spin-axis direction $\hat{\mathbf{n}}$. In the case of two sublattice collinear AFM, $\hat{\mathbf{n}} = \Lbhat = \Lbhat /|\mathbf{L}|$, where $\mathbf{L}$ is the N\'eel vector: $\mathbf{L} = \mathbf{M}_A - \mathbf{M}_B$. In FMs $\hat{\mathbf{n}} = \mathbf{M}/| \mathbf{M}|$. We can then write the tensor $\chi_a$ in the following way, \cite{Hals2013}
\begin{align}
 \chi_{a,ij}(\hat{\mathbf{n}}) = \chi_{a,ij}^{(0)} + \chi_{a,ij,k}^{(1)} \hat{n}_k + \chi_{a,ij,kl}^{(2)} \hat{n}_k \hat{n}_l + \dots
 \label{eq:chi_expansion}
\end{align}
Here the Einstein summation notation is used. Note that since $\hat{\mathbf{n}}$ is a unit vector, the expansion could be done using two variables only. We find it more practical, however, to use all three components of $\hat{\mathbf{n}}$. The odd terms in the expansion correspond to the odd part of the CISP, while the even terms correspond to the even part.

To find the symmetry properties of the expansion \eqref{eq:chi_expansion} we have to consider the nonmagnetic local point group. This is a group of symmetry operations of the nonmagnetic crystal that leave the sublattice $a$ invariant. (See the Appendix \ref{appendix:symmetry} for details on how to find the symmetry properties of the expansion \eqref{eq:chi_expansion}.) Since there are only 21 nonmagnetic point groups with broken inversion symmetry, it is feasible to calculate all allowed leading terms of the expansion \eqref{eq:chi_expansion}. This was done  for the zeroth order terms in Ref. \cite{ciccarelli2015} that focused on the CISP in FMs. The zeroth order terms generate the field-like torque. In  Table \ref{table:point_groups1} we give all allowed first order terms and for completeness we also show the zeroth order terms. The zeroth order term vanishes for several point groups. For those we also give the second order terms in Table \ref{table:point_groups2}. Together the tables give the lowest order terms for the even and odd part of the CISP in all 21 non-centrosymmetric point groups.

The tensors in Tables \ref{table:point_groups1},\ref{table:point_groups2} are given in cartesian coordinate systems. The cartesian systems are defined in terms of the conventional basis vectors $\mathbf{a}, \mathbf{b}, \mathbf{c}$ (see the International Tables for Crystallography \cite{crystallographictables}). The choice of the cartesian system is straightforward for the orthorhombic, tetragonal and cubic groups. The tensors for the triclinic group $1$ have a completely general form and the choice of the coordinate system is thus irrelevant for this group.  For hexagonal and trigonal groups, we choose the right-handed coordinate system that satisfies $\mathbf{x} = \mathbf{a}/|\mathbf{a}|$, $\mathbf{z} = \mathbf{c}/| \mathbf{c}|$. For the monoclinic groups we use the unique axis $b$ setting \cite{crystallographictables} and choose the right-handed coordinate system that satisfies $\mathbf{x} = \mathbf{a}/|\mathbf{a}|$, $\mathbf{y} = \mathbf{b}/|\mathbf{b}|$.

The tensors in Tables \ref{table:point_groups1},\ref{table:point_groups2} apply for two-sublattice collinear AFMs and FMs. In the case of AFMs the expansion only applies for the CISP on a sublattice and correspondingly the local point group has to be used. In FMs, the tensors apply for the local as well as for the net CISP. In the latter case the global point group has to be used. Since the zeroth order term is independent of magnetic moments it can be equally considered for any material, including non-collinear AFMs. In nonmagnetic materials, there is naturally  no dependence on magnetic moments so the zeroth order terms describes the CISP completely in this case.

The zeroth order terms that generate the field-like torque are particularly important since they are often dominant. As discussed in Ref. \cite{ciccarelli2015}, the tensors corresponding to the field-like torque are in general composed of three distinct terms: generalized Rashba and Dresselhaus terms and a term describing a response proportional to the electric field. They are described by the following tensors respectively
\begin{align}
 \chi_a^{\text{gR}} &= \left(\begin{matrix}x_{11} & - x_{21} & 0\\x_{21} & x_{11} & 0\\0 & 0 & 0\end{matrix}\right), \label{eq:g_Rashba} \\
 \chi_a^{\text{gD}} &= \left(\begin{matrix}x_{11} & x_{21} & 0\\x_{21} & - x_{11} & 0\\0 & 0 & 0\end{matrix}\right), \label{eq:g_Dresselhaus}\\
 \chi_a^{E} &= \left(\begin{matrix}x_{11} & 0 & 0\\0 & x_{11} & 0\\0 & 0 & x_{11}\end{matrix}\right).\label{eq:chi_E}
\end{align}

\begin{figure}[h]
  \includegraphics[width=8.6cm]{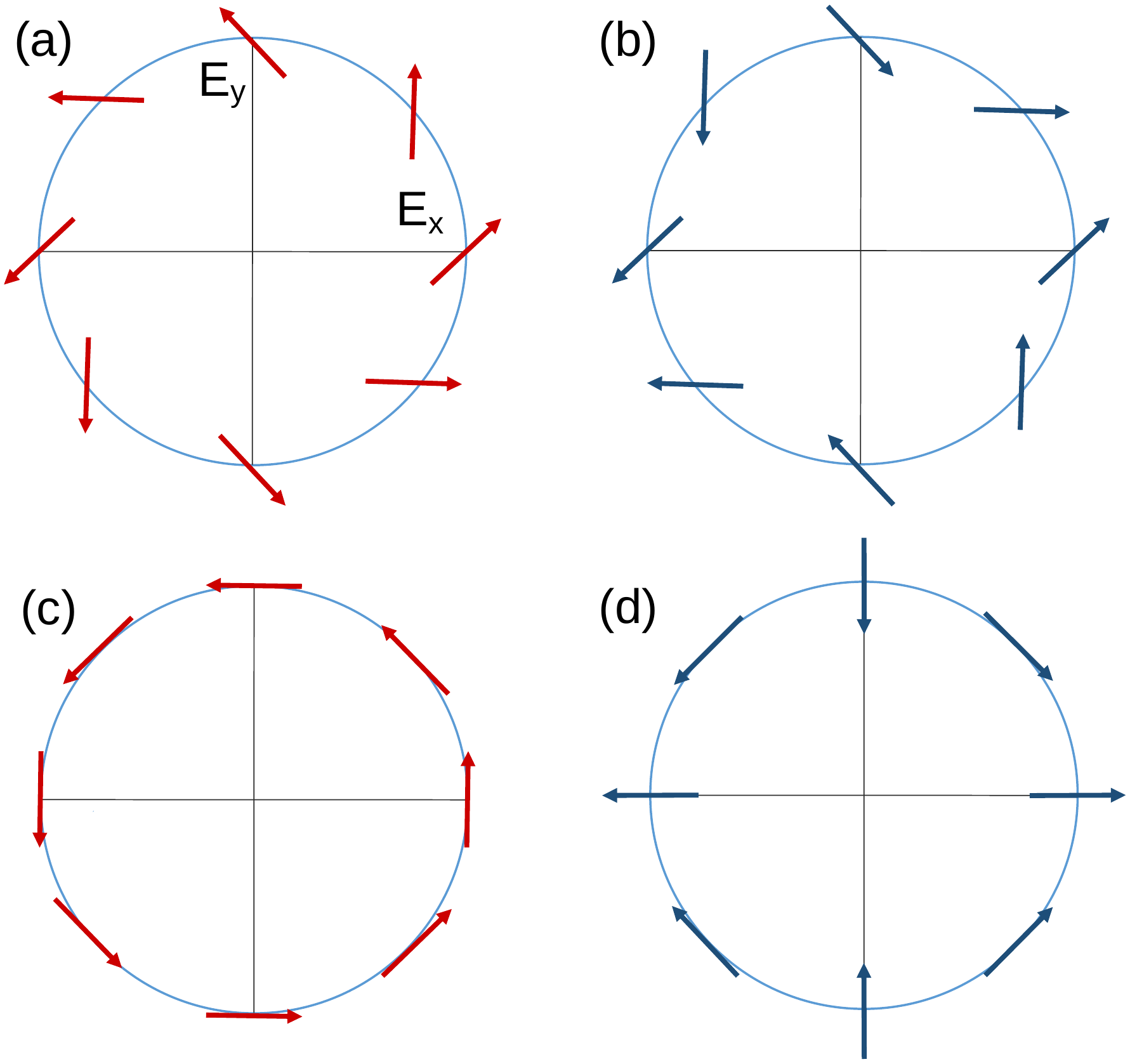}
  \caption{(Color online) Illustration of the Rashba and Dresselhaus CISPs. The figures show the dependence of CISP on the electric field direction. Adapted from \cite{ciccarelli2015}. (a) Generalized Rashba CISP, (b) generalized Dresselhaus CISP, (c) Rashba CISP, (d) Dresselhaus CISP.}
  \label{fig:rasba_dresselhaus}
\end{figure}

The generalized Rashba and Dresselhaus CISPs lie in a plane and are only present for the current applied in the same plane. In Eqs. \eqref{eq:g_Rashba}, \eqref{eq:g_Dresselhaus}, it is the $xy$ plane, but in general, it can be any plane. The generalized Rashba and Dresselhaus terms differ in how the CISP depends on the current direction, as illustrated in Figs \ref{fig:rasba_dresselhaus}(a),(b). In the case of the Rashba CISP, when the current direction is rotated the CISP rotates in the same way, while in the case of the Dresselhaus CISP the field rotates in the opposite direction. They differ from the conventional Rashba and Dresselhaus terms,
\begin{align}
 \chi_a^{\text{R}} &= \left(\begin{matrix} 0 & - x_{21} & 0\\x_{21} & 0& 0\\0 & 0 & 0\end{matrix}\right), \\
 \chi_a^{\text{D}} &= \left(\begin{matrix}x_{11} & 0 & 0\\0 & - x_{11} & 0\\0 & 0 & 0\end{matrix}\right),
\end{align}
by a constant off-set angle between the applied current and the CISP (see Figs. \ref{fig:rasba_dresselhaus}(c),(d)).

The Rashba term or generalized Rashba term with nonzero $x_{21}$ components occurs in polar point groups (groups $1$, $2$, $m$, $mm2$, $4$, $4mm$, $3$, $3m1$, $6$ and $6mm$), i.e., in groups which allow the existence of a permanent electric dipole moment. In a polar group, there is a Rashba term in the plane perpendicular to the electric dipole moment. The Rashba term can be written as $\delta \mathbf{S} \sim \hat{\mathbf{D}} \times \mathbf{E}$,  where $\hat{\mathbf{D}}$ is the direction of the electric dipole moment. In all polar groups except for $m$ and $1$, $\hat{\mathbf{D}}$ is oriented along the polar direction (a direction invariant under all symmetry operations), which in the coordinate systems used in Table \ref{table:point_groups1} is always oriented along the $\mathbf{z}$ axis. In group $1$, $\hat{\mathbf{D}}$ can have any direction, and in the group $m$, it is oriented in the mirror plane.  Polar point groups $1$, $2$, $4$, $3$ and $6$ contain the generalized Rashba term (rather than just the Rashba term), which in addition also occurs in nonpolar point groups $222$, $422$, $312$ and $622$ with $x_{21} = 0$.  The CISP described by $\chi_a^E$ in Eq. \eqref{eq:chi_E} occurs in  enantiomorphic (also called chiral) crystals (point groups $1$, $2$, $222$, $4$, $422$, $3$, $312$, $6$, $622$, $23$ and $432$), i.e., crystals in which no symmetry operation contains inversion. The CISP is an axial vector (even under inversion), while the electric field is a polar vector (odd under inversion). These two vectors can only be proportional in the enantiomorphic crystals since in these crystals there is no difference between an axial and polar vector. The generalized Dresselhaus term (Eq. \eqref{eq:g_Dresselhaus}) occurs in point groups $1$, $2$, $m$, $222$, $mm2$, $-4$ and $-42m$, of which the groups $222$ and $-42m$ have just the Dresselhaus term.

The local nonmagnetic point group of both the 3D $\text{Mn}_2\text{Au}$ and the 2D Rashba model is $4mm$, which has a Rashba zeroth order CISP of the form $\mathbf{z} \times \mathbf{E}$. Another example of an AFM with Rashba zeroth order CISP is CuMnAs \cite{Wadley2016}. Dresselhaus zeroth order CISPs have been previously observed in FMs GaMnAs \cite{Fang2011,Kurebayashi2014} and NiMnSb \cite{ciccarelli2015} (global point group $-42m$). AFM CuMnSb (local point group $-42m$ \cite{Forster1968}) is another example for which we expect the Dresselhaus zeroth order CISP, according to our symmetry analysis.

The first order term for the $4mm$ point group can be written in the following way
\begin{align}
 \chi_a^{(1)} =  X_1 + X_2 + X_3,\label{eq:exp_odd}
\end{align}
where
\begin{align}
X_1 &=   C_1\left( \begin{matrix} 		-\hat{L}_z & 0 & 0  \\
					0 & -\hat{L}_z & 0 \\
					\hat{L}_x & \hat{L}_y & 0
                                       \end{matrix}\right),\\
X_2 &=  C_2 \left( \begin{matrix} 		0 & 0 & 0  \\
					0 & 0 & 0 \\
					\hat{L}_x & \hat{L}_y & 0
                                       \end{matrix}\right),\\
X_3 &=  \left( \begin{matrix}
			0 & 0 & C_3 \hat{L}_x \\
			0 & 0 & C_3 \hat{L}_y \\
			0 & 0 & C_4 \hat{L}_z
			\end{matrix}\right).
\end{align}
Note that in the 2D Rashba model the current cannot flow in the $z$ direction, so the third column has no physical meaning in this case. The contribution to the CISP generated by the tensor $X_1$ can be constructed from the polar direction of the group $4mm$: $\Lbhat \times (\mathbf{z} \times \mathbf{E})$. The contribution to the CISP coming from the tensor $X_2$ can be written as  $(\Lbhat \cdot \mathbf{E}_\parallel ) \mathbf{z}$, where $\mathbf{E}_\parallel$ is the in-plane (the plane here refers to the $xy$ plane) component of the electric field. Finally we note the connection of our general symmetry analysis of the CISP to the discussion in Ref. \cite{Zhang2014} of Rashba and Dresselhaus-like spin-orbit coupling effects on the level of equilibrium electronic structure of locally non-centrosymmetric crystals.
\begingroup
\squeezetable
\begin{longtable*}{cccc}
 \hline \hline 
 \rule{0pt}{11pt}   
 Crystal system & Point group & $\chi^{(0)}$ & $\chi^{(1)}$ \\[1pt]
 \hline
 \rule{0pt}{17pt}   
 triclinic & 1 &  $ \left(\begin{matrix}x_{11} & x_{12} & x_{13}\\x_{21} & x_{22} & x_{23}\\x_{31} & x_{32} & x_{33}\end{matrix}\right)
$ & $ \left(\begin{matrix}\hat{n}_{x} x_{111} + \hat{n}_{y} x_{112} + \hat{n}_{z} x_{113} & \hat{n}_{x} x_{121} + \hat{n}_{y} x_{122} + \hat{n}_{z} x_{123} & \hat{n}_{x} x_{131} + \hat{n}_{y} x_{132} + \hat{n}_{z} x_{133}\\\hat{n}_{x} x_{211} + \hat{n}_{y} x_{212} + \hat{n}_{z} x_{213} & \hat{n}_{x} x_{221} + \hat{n}_{y} x_{222} + \hat{n}_{z} x_{223} & \hat{n}_{x} x_{231} + \hat{n}_{y} x_{232} + \hat{n}_{z} x_{233}\\\hat{n}_{x} x_{311} + \hat{n}_{y} x_{312} + \hat{n}_{z} x_{313} & \hat{n}_{x} x_{321} + \hat{n}_{y} x_{322} + \hat{n}_{z} x_{323} & \hat{n}_{x} x_{331} + \hat{n}_{y} x_{332} + \hat{n}_{z} x_{333}\end{matrix}\right)
 $ \\[10pt]
monoclinic & 2 &  $ \left(\begin{matrix}x_{11} & 0 & x_{13}\\0 & x_{22} & 0\\x_{31} & 0 & x_{33}\end{matrix}\right)
$ & $ \left(\begin{matrix}\hat{n}_{y} x_{1} & \hat{n}_{x} x_{13} + \hat{n}_{z} x_{12} & \hat{n}_{y} x_{3}\\\hat{n}_{x} x_{5} + \hat{n}_{z} x_{6} & \hat{n}_{y} x_{11} & \hat{n}_{x} x_{4} + \hat{n}_{z} x_{7}\\\hat{n}_{y} x_{10} & \hat{n}_{x} x_{8} + \hat{n}_{z} x_{9} & \hat{n}_{y} x_{2}\end{matrix}\right)
 $ \\[10pt]
& m &  $ \left(\begin{matrix}0 & x_{12} & 0\\x_{21} & 0 & x_{23}\\0 & x_{32} & 0\end{matrix}\right)
$ & $ \left(\begin{matrix}\hat{n}_{x} x_{12} + \hat{n}_{z} x_{9} & \hat{n}_{y} x_{14} & \hat{n}_{x} x_{13} + \hat{n}_{z} x_{8}\\\hat{n}_{y} x_{3} & \hat{n}_{x} x_{11} + \hat{n}_{z} x_{10} & \hat{n}_{y} x_{4}\\\hat{n}_{x} x_{7} + \hat{n}_{z} x_{6} & \hat{n}_{y} x_{5} & \hat{n}_{x} x_{1} + \hat{n}_{z} x_{2}\end{matrix}\right)
 $ \\[10pt]
orthorhombic & 222 &  $ \left(\begin{matrix}x_{11} & 0 & 0\\0 & x_{22} & 0\\0 & 0 & x_{33}\end{matrix}\right)
$ & $ \left(\begin{matrix}0 & \hat{n}_{z} x_{5} & \hat{n}_{y} x_{4}\\\hat{n}_{z} x_{1} & 0 & \hat{n}_{x} x_{6}\\\hat{n}_{y} x_{3} & \hat{n}_{x} x_{2} & 0\end{matrix}\right)
 $ \\[10pt]
& mm2 &  $ \left(\begin{matrix}0 & x_{12} & 0\\x_{21} & 0 & 0\\0 & 0 & 0\end{matrix}\right)
$ & $ \left(\begin{matrix}\hat{n}_{z} x_{4} & 0 & \hat{n}_{x} x_{6}\\0 & \hat{n}_{z} x_{5} & \hat{n}_{y} x_{7}\\\hat{n}_{x} x_{3} & \hat{n}_{y} x_{2} & \hat{n}_{z} x_{1}\end{matrix}\right)
 $ \\[10pt]
tetragonal & 4 &  $ \left(\begin{matrix}x_{11} & - x_{21} & 0\\x_{21} & x_{11} & 0\\0 & 0 & x_{33}\end{matrix}\right)
$ & $ \left(\begin{matrix}\hat{n}_{z} x_{6} & - \hat{n}_{z} x_{2} & \hat{n}_{x} x_{5} - \hat{n}_{y} x_{7}\\\hat{n}_{z} x_{2} & \hat{n}_{z} x_{6} & \hat{n}_{x} x_{7} + \hat{n}_{y} x_{5}\\\hat{n}_{x} x_{4} - \hat{n}_{y} x_{3} & \hat{n}_{x} x_{3} + \hat{n}_{y} x_{4} & \hat{n}_{z} x_{1}\end{matrix}\right)
 $ \\[10pt]
& -4 &  $ \left(\begin{matrix}x_{11} & x_{21} & 0\\x_{21} & - x_{11} & 0\\0 & 0 & 0\end{matrix}\right)
$ & $ \left(\begin{matrix}\hat{n}_{z} x_{5} & \hat{n}_{z} x_{1} & \hat{n}_{x} x_{4} + \hat{n}_{y} x_{6}\\\hat{n}_{z} x_{1} & - \hat{n}_{z} x_{5} & \hat{n}_{x} x_{6} - \hat{n}_{y} x_{4}\\\hat{n}_{x} x_{3} + \hat{n}_{y} x_{2} & \hat{n}_{x} x_{2} - \hat{n}_{y} x_{3} & 0\end{matrix}\right)
 $ \\[10pt]
& 422 &  $ \left(\begin{matrix}x_{11} & 0 & 0\\0 & x_{11} & 0\\0 & 0 & x_{33}\end{matrix}\right)
$ & $ \left(\begin{matrix}0 & - \hat{n}_{z} x_{3} & - \hat{n}_{y} x_{2}\\\hat{n}_{z} x_{3} & 0 & \hat{n}_{x} x_{2}\\- \hat{n}_{y} x_{1} & \hat{n}_{x} x_{1} & 0\end{matrix}\right)
 $ \\[10pt]
& 4mm &  $ \left(\begin{matrix}0 & - x_{21} & 0\\x_{21} & 0 & 0\\0 & 0 & 0\end{matrix}\right)
$ & $ \left(\begin{matrix}\hat{n}_{z} x_{4} & 0 & \hat{n}_{x} x_{1}\\0 & \hat{n}_{z} x_{4} & \hat{n}_{y} x_{1}\\\hat{n}_{x} x_{3} & \hat{n}_{y} x_{3} & \hat{n}_{z} x_{2}\end{matrix}\right)
 $ \\[10pt]
& -42m &  $ \left(\begin{matrix}x_{11} & 0 & 0\\0 & - x_{11} & 0\\0 & 0 & 0\end{matrix}\right)
$ & $ \left(\begin{matrix}0 & \hat{n}_{z} x_{3} & \hat{n}_{y} x_{2}\\\hat{n}_{z} x_{3} & 0 & \hat{n}_{x} x_{2}\\\hat{n}_{y} x_{1} & \hat{n}_{x} x_{1} & 0\end{matrix}\right)
 $ \\[10pt]
& -4m2 &  $ \left(\begin{matrix}0 & x_{21} & 0\\x_{21} & 0 & 0\\0 & 0 & 0\end{matrix}\right)
$ & $ \left(\begin{matrix}\hat{n}_{z} x_{3} & 0 & \hat{n}_{x} x_{1}\\0 & - \hat{n}_{z} x_{3} & - \hat{n}_{y} x_{1}\\\hat{n}_{x} x_{2} & - \hat{n}_{y} x_{2} & 0\end{matrix}\right)
 $ \\[10pt]
trigonal & 3 &  $ \left(\begin{matrix}x_{11} & - x_{21} & 0\\x_{21} & x_{11} & 0\\0 & 0 & x_{33}\end{matrix}\right)
$ & $ \left(\begin{matrix}\hat{n}_{x} x_{7} + \hat{n}_{y} x_{2} + \hat{n}_{z} x_{8} & \hat{n}_{x} x_{2} - \hat{n}_{y} x_{7} - \hat{n}_{z} x_{3} & \hat{n}_{x} x_{6} - \hat{n}_{y} x_{9}\\\hat{n}_{x} x_{2} - \hat{n}_{y} x_{7} + \hat{n}_{z} x_{3} & - \hat{n}_{x} x_{7} - \hat{n}_{y} x_{2} + \hat{n}_{z} x_{8} & \hat{n}_{x} x_{9} + \hat{n}_{y} x_{6}\\\hat{n}_{x} x_{5} - \hat{n}_{y} x_{4} & \hat{n}_{x} x_{4} + \hat{n}_{y} x_{5} & \hat{n}_{z} x_{1}\end{matrix}\right)
 $ \\[10pt]
& 312 &  $ \left(\begin{matrix}x_{11} & 0 & 0\\0 & x_{11} & 0\\0 & 0 & x_{33}\end{matrix}\right)
$ & $ \left(\begin{matrix}\hat{n}_{y} x_{3} & \hat{n}_{x} x_{3} - \hat{n}_{z} x_{4} & - \hat{n}_{y} x_{2}\\\hat{n}_{x} x_{3} + \hat{n}_{z} x_{4} & - \hat{n}_{y} x_{3} & \hat{n}_{x} x_{2}\\- \hat{n}_{y} x_{1} & \hat{n}_{x} x_{1} & 0\end{matrix}\right)
 $ \\[10pt]
& 321 &  $ \left(\begin{matrix}x_{11} & 0 & 0\\0 & x_{11} & 0\\0 & 0 & x_{33}\end{matrix}\right)
$ & $ \left(\begin{matrix}\hat{n}_{x} x_{3} & - \hat{n}_{y} x_{3} - \hat{n}_{z} x_{4} & - \hat{n}_{y} x_{2}\\- \hat{n}_{y} x_{3} + \hat{n}_{z} x_{4} & - \hat{n}_{x} x_{3} & \hat{n}_{x} x_{2}\\- \hat{n}_{y} x_{1} & \hat{n}_{x} x_{1} & 0\end{matrix}\right)
 $ \\[10pt]
& 3m1 &  $ \left(\begin{matrix}0 & - x_{21} & 0\\x_{21} & 0 & 0\\0 & 0 & 0\end{matrix}\right)
$ & $ \left(\begin{matrix}\hat{n}_{y} x_{4} + \hat{n}_{z} x_{5} & \hat{n}_{x} x_{4} & \hat{n}_{x} x_{2}\\\hat{n}_{x} x_{4} & - \hat{n}_{y} x_{4} + \hat{n}_{z} x_{5} & \hat{n}_{y} x_{2}\\\hat{n}_{x} x_{3} & \hat{n}_{y} x_{3} & \hat{n}_{z} x_{1}\end{matrix}\right)
 $ \\[10pt]
& 31m &  $ \left(\begin{matrix}0 & - x_{21} & 0\\x_{21} & 0 & 0\\0 & 0 & 0\end{matrix}\right)
$ & $ \left(\begin{matrix}\hat{n}_{x} x_{3} + \hat{n}_{z} x_{5} & - \hat{n}_{y} x_{3} & \hat{n}_{x} x_{1}\\- \hat{n}_{y} x_{3} & - \hat{n}_{x} x_{3} + \hat{n}_{z} x_{5} & \hat{n}_{y} x_{1}\\\hat{n}_{x} x_{4} & \hat{n}_{y} x_{4} & \hat{n}_{z} x_{2}\end{matrix}\right)
 $ \\[10pt]
hexagonal & 6 &  $ \left(\begin{matrix}x_{11} & - x_{21} & 0\\x_{21} & x_{11} & 0\\0 & 0 & x_{33}\end{matrix}\right)
$ & $ \left(\begin{matrix}\hat{n}_{z} x_{6} & - \hat{n}_{z} x_{2} & \hat{n}_{x} x_{5} - \hat{n}_{y} x_{7}\\\hat{n}_{z} x_{2} & \hat{n}_{z} x_{6} & \hat{n}_{x} x_{7} + \hat{n}_{y} x_{5}\\\hat{n}_{x} x_{4} - \hat{n}_{y} x_{3} & \hat{n}_{x} x_{3} + \hat{n}_{y} x_{4} & \hat{n}_{z} x_{1}\end{matrix}\right)
 $ \\[10pt]
& -6 &  $ \left(\begin{matrix}0 & 0 & 0\\0 & 0 & 0\\0 & 0 & 0\end{matrix}\right)
$ & $ \left(\begin{matrix}\hat{n}_{x} x_{1} + \hat{n}_{y} x_{2} & \hat{n}_{x} x_{2} - \hat{n}_{y} x_{1} & 0\\\hat{n}_{x} x_{2} - \hat{n}_{y} x_{1} & - \hat{n}_{x} x_{1} - \hat{n}_{y} x_{2} & 0\\0 & 0 & 0\end{matrix}\right)
 $ \\[10pt]
& 622 &  $ \left(\begin{matrix}x_{11} & 0 & 0\\0 & x_{11} & 0\\0 & 0 & x_{33}\end{matrix}\right)
$ & $ \left(\begin{matrix}0 & - \hat{n}_{z} x_{3} & - \hat{n}_{y} x_{2}\\\hat{n}_{z} x_{3} & 0 & \hat{n}_{x} x_{2}\\- \hat{n}_{y} x_{1} & \hat{n}_{x} x_{1} & 0\end{matrix}\right)
 $ \\[10pt]
& 6mm &  $ \left(\begin{matrix}0 & - x_{21} & 0\\x_{21} & 0 & 0\\0 & 0 & 0\end{matrix}\right)
$ & $ \left(\begin{matrix}\hat{n}_{z} x_{4} & 0 & \hat{n}_{x} x_{1}\\0 & \hat{n}_{z} x_{4} & \hat{n}_{y} x_{1}\\\hat{n}_{x} x_{3} & \hat{n}_{y} x_{3} & \hat{n}_{z} x_{2}\end{matrix}\right)
 $ \\[10pt]
& -6m2 &  $ \left(\begin{matrix}0 & 0 & 0\\0 & 0 & 0\\0 & 0 & 0\end{matrix}\right)
$ & $ \left(\begin{matrix}\hat{n}_{y} x_{1} & \hat{n}_{x} x_{1} & 0\\\hat{n}_{x} x_{1} & - \hat{n}_{y} x_{1} & 0\\0 & 0 & 0\end{matrix}\right)
 $ \\[10pt]
& -62m &  $ \left(\begin{matrix}0 & 0 & 0\\0 & 0 & 0\\0 & 0 & 0\end{matrix}\right)
$ & $ \left(\begin{matrix}\hat{n}_{x} x_{1} & - \hat{n}_{y} x_{1} & 0\\- \hat{n}_{y} x_{1} & - \hat{n}_{x} x_{1} & 0\\0 & 0 & 0\end{matrix}\right)
 $ \\[10pt]
cubic & 23 &  $ \left(\begin{matrix}x_{11} & 0 & 0\\0 & x_{11} & 0\\0 & 0 & x_{11}\end{matrix}\right)
$ & $ \left(\begin{matrix}0 & \hat{n}_{z} x_{2} & \hat{n}_{y} x_{1}\\\hat{n}_{z} x_{1} & 0 & \hat{n}_{x} x_{2}\\\hat{n}_{y} x_{2} & \hat{n}_{x} x_{1} & 0\end{matrix}\right)
 $ \\[10pt]
& 432 &  $ \left(\begin{matrix}x_{11} & 0 & 0\\0 & x_{11} & 0\\0 & 0 & x_{11}\end{matrix}\right)
$ & $ \left(\begin{matrix}0 & - \hat{n}_{z} x_{1} & \hat{n}_{y} x_{1}\\\hat{n}_{z} x_{1} & 0 & - \hat{n}_{x} x_{1}\\- \hat{n}_{y} x_{1} & \hat{n}_{x} x_{1} & 0\end{matrix}\right)
 $ \\[10pt]
& -43m &  $ \left(\begin{matrix}0 & 0 & 0\\0 & 0 & 0\\0 & 0 & 0\end{matrix}\right)
$ & $ \left(\begin{matrix}0 & \hat{n}_{z} x_{1} & \hat{n}_{y} x_{1}\\\hat{n}_{z} x_{1} & 0 & \hat{n}_{x} x_{1}\\\hat{n}_{y} x_{1} & \hat{n}_{x} x_{1} & 0\end{matrix}\right)
 $ \\[10pt]
\hline \hline
\noalign{\vskip 1mm}  
\caption{Zeroth and first order terms in the expansion \eqref{eq:chi_expansion} for the point groups with broken inversion symmetry. The tensors $\chi^{(1)}$ have the spin-axis direction included: $\chi^{(1)}_{ij} = \chi^{(1)}_{ij,k}\hat{n}_k$. The $x$ parameters can be chosen arbitrarily for each tensor. Note that the groups $-42m$ and $-4m2$, $312$ and $321$, $3m1$ and $31m$, and $-6m2$ and $-62m$ are equivalent and differ only by a coordinate transformation. For completeness we also give the tensors for the equivalent groups.}
\label{table:point_groups1}
\end{longtable*}
\endgroup

\begingroup
\squeezetable
\begin{longtable*}{cc}
\hline \hline
\rule{0pt}{11pt}   
point group & $\chi^{(2)}$ \\
\hline   
\rule{0pt}{17pt}   
-6 & $\left(\begin{matrix}\hat{n}_{z} \left(\hat{n}_{x} x_{5}  + \hat{n}_{y} x_{4} \right) & \hat{n}_{z} \left(\hat{n}_{x} x_{4} - \hat{n}_{y} x_{5}\right) & \hat{n}_{x}^{2} x_{2} + 2 \hat{n}_{x} \hat{n}_{y} x_0 - \hat{n}_{y}^{2} x_{2}\\\hat{n}_{z} \left(\hat{n}_{x} x_{4} - \hat{n}_{y} x_{5} \right) & - \hat{n}_{z} \left(\hat{n}_{x} x_{5} + \hat{n}_{y} x_{4} \right) & \hat{n}_{x}^{2} x_0 - 2 \hat{n}_{x} \hat{n}_{y} x_{2} - \hat{n}_{y}^{2} x_0\\\hat{n}_{x}^{2} x_{1} + 2 \hat{n}_{x} \hat{n}_{y} x_{3} - \hat{n}_{y}^{2} x_{1} & \hat{n}_{x}^{2} x_{3} - 2 \hat{n}_{x} \hat{n}_{y} x_{1} - \hat{n}_{y}^{2} x_{3} & 0\end{matrix}\right) $ \\[15pt]
-6m2 & $\left(\begin{matrix}\hat{n}_{x} \hat{n}_{z} x_1 & - \hat{n}_{y} \hat{n}_{z} x_1 & x_{2} \left(\hat{n}_{x}^{2} - \hat{n}_{y}^{2}\right)\\- \hat{n}_{y} \hat{n}_{z} x_1 & - \hat{n}_{x} \hat{n}_{z} x_1 & - 2 \hat{n}_{x} \hat{n}_{y} x_{2}\\x_3 \left(\hat{n}_{x}^{2} - \hat{n}_{y}^{2}\right) & - 2 \hat{n}_{x} \hat{n}_{y} x_3 & 0\end{matrix}\right)$ \\[20pt]
 -62m &  $ \left(\begin{matrix}\hat{n}_{y} \hat{n}_{z} x_1 & \hat{n}_{x} \hat{n}_{z} x_1 & 2 \hat{n}_{x} \hat{n}_{y} x_{2}\\\hat{n}_{x} \hat{n}_{z} x_1 & - \hat{n}_{y} \hat{n}_{z} x_1 & x_{2} \left(\hat{n}_{x}^{2} - \hat{n}_{y}^{2}\right)\\2 \hat{n}_{x} \hat{n}_{y} x_3 & x_3 \left(\hat{n}_{x}^{2} - \hat{n}_{y}^{2}\right) & 0\end{matrix}\right) $ \\[15pt]
-43m & $\left(\begin{matrix}x_{2} \left(- \hat{n}_{y}^{2} + \hat{n}_{z}^{2}\right) & \hat{n}_{x} \hat{n}_{y}x_1 & -\hat{n}_{x} \hat{n}_{z} x_1 \\-\hat{n}_{x} \hat{n}_{y} x_1 & x_{2} \left(\hat{n}_{x}^{2} - \hat{n}_{z}^{2}\right) & \hat{n}_{y} \hat{n}_{z} x_1\\\hat{n}_{x} \hat{n}_{z} x_1 & -\hat{n}_{y} \hat{n}_{z} x_1& x_{2} \left(- \hat{n}_{x}^{2} + \hat{n}_{y}^{2}\right)\end{matrix}\right)$ \\[10pt]
\hline \hline
\noalign{\vskip 1mm}  
\caption{Second order terms in the expansion \eqref{eq:chi_expansion} for the point groups which have no zeroth order term allowed by symmetry. The $x$ parameters can be chosen arbitrarily for each tensor.}
\label{table:point_groups2}
\end{longtable*}
\endgroup

\section{Linear response theory}
\label{sec:results}
\subsection{Analytical calculations}
\label{sec:analytical}
The CISP in the AFM 2D Rashba model can be calculated analytically when magnetic moments are oriented close to the out-of-plane ($\mathbf{z}$) direction and when the Fermi level is close to the bottom or the top of the bands so that only $\kb$ points close to the $\Gamma$ point matter for the torque calculation. To the second order in $k$ the Hamiltonian \eqref{eq:2Dmodel} can be expressed as
\begin{eqnarray}\label{eq:hfk}
\hat{H}=&\gamma_k\hat\tau_x-\alpha k\hat{\bm\sigma}\cdot{\bm\mu}\hat\tau_x+J_{\rm sd}\Lbhat\cdot\hat{\bm\sigma}\hat\tau_z,\label{eq:ham_aurelien}
\end{eqnarray}
where $\gamma_k = ta_l^2\left(k^2-(2/a_l)^2\right)$, ${\bm\mu}$ is a unit vector perpendicular to the $\mathbf{k}$ vector, expressed as  ${\bm \mu}=(\sin\varphi_k,-\cos\varphi_k,0)$, where $\varphi_k$ is defined by ${\bf k}=k(\cos\varphi_k,\sin\varphi_k,0)$. $\hat{\sigmab}$ and $\hat{\bm\tau}$ are Pauli matrices with  $\hat{\sigmab}/2$ representing the carrier spin degree of freedom and $\hat{\bm\tau}$ the AFM spin-sublattice degree of freedom of carriers.

Hamiltonian \eqref{eq:ham_aurelien} can be solved analytically \cite{Saidaouiunpublished}. The associated {\em unperturbed} retarded Green's function, defined as $\hat{G}^R_0=(\epsilon-\hat{H}+i0^+)^{-1}$, reads
\begin{eqnarray}\label{eq:go}
\hat{G}^R_{0}&=&\frac{1}{4S_k}\sum_{s,\eta=\pm1}\frac{1}{\epsilon-\epsilon_{s,\eta}+i0^+}\times\\
&&\left[S_k+s(\gamma_k(\hat{\bm\sigma}\cdot{\bm\mu})-J_{\rm sd}\hat{\tau}_y(\hat{\bm\sigma}\cdot\Lbhat\times{\bm\mu}))\right.\nonumber\\
&&\left.+\frac{1}{\epsilon_{s,\eta}}\left((s\gamma_k^2+sJ_{\rm sd}^2+\alpha k S_k)(\hat{\bm\sigma}\cdot{\bm\mu})\hat{\tau}_x\right.\right.\nonumber\\
&&\left.\left.+(\gamma_k\hat{\tau}_x+J_{\rm sd}(\hat{\bm\sigma}\cdot\Lbhat)\hat{\tau}_z)(S_k+s\alpha k)\right.\right.\nonumber\\
&&\left.\left.-sJ_{\rm sd}(\Lbhat\cdot{\bm\mu})(J_{\rm sd}(\hat{\bm\sigma}\cdot\Lbhat)\hat{\tau}_x-\gamma_k\hat{\tau}_z+\alpha k(\hat{\bm\sigma}\cdot{\bm\mu})\hat{\tau}_z)\right)\right]\nonumber,
\end{eqnarray}
where $\epsilon_{s,\eta}$ denotes the band structure given by
\begin{eqnarray}
\epsilon_{s,\eta}&=&\eta\sqrt{\gamma_k^2+J_{\rm sd}^2+\alpha^2k^2+2s\alpha kS_k},\\
S_k&=&\sqrt{\gamma_k^2+J_{\rm sd}^2(1-\sin^2\theta\sin^2(\varphi_k-\varphi))}.\label{eq:sk}
\end{eqnarray}
Indices $s,\eta$ refer to the spin chirality ($s=\pm1$) and to the electron/hole bands ($\eta=\pm1$). In the limit of vanishing $\alpha$, both spin chiralities become degenerate. The angles $\theta$, $\varphi$ are spherical coordinates of the vector $\Lbhat$. In order to get an analytically tractable expression for the CISP, we express in the following  the Green's function in term of the projection operator ${\cal A}_{s,\eta}=|s,\eta\rangle\langle s,\eta|$ such that $\hat{G}^{R}_{0}=\sum_{s,\eta}{\cal A}_{s,\eta}/(\epsilon-\epsilon_{s,\eta}+ i0^+)$. 

We evaluate the intraband term using the expression \eqref{Boltzmann}, which applies for small $\Gamma$. For the interband term we take the $\Gamma \rightarrow 0$ limit in which the term $\chi_{a}^\text{II(b)}$ is zero and the term $\chi_{a}^\text{II(a)}$ is constant. 
The CISP can then be written as 
\begin{align}\label{eq:intra}
{\bf S}^{\rm Intra}=&\frac{e\hbar}{2\Gamma V}\sum_{\nu,{\bf k}}{\rm Re}\{{\rm Tr}[(\hat{\bm v}\cdot{\bf E}){\cal A}_\nu{\bm\sigma}_\varsigma{\cal A}_\nu]\}\delta(\epsilon_{{\bf k},\nu}-\epsilon_{\rm F}),\\
%{\bf S}^{\rm Inter1}=&&-\frac{2e\hbar\Gamma}{V}\sum_{\nu\neq \nu',{\bf k}}{\rm Re} \{{\rm Tr}[(\hat{\bm v}\cdot{\bf E}){\cal A}_\nu{\bm\sigma}_\varsigma{\cal A}_{\nu'}]\}\frac{(f_{{\bf k},\nu}-f_{{\bf k},\nu'})}{(\epsilon_{{\bf k},\nu}-\epsilon_{{\bf k},\nu'})^3},\label{eq:inter1}\\
{\bf S}^{\rm Inter}=&\frac{e\hbar}{V}\sum_{\nu\neq \nu',{\bf k}}{\rm Im}{\rm Tr}[(\hat{\bm v}\cdot{\bf E}){\cal A}_\nu{\bm\sigma}_\varsigma{\cal A}_{\nu'}]\}\frac{(f_{{\bf k},\nu}-f_{{\bf k},\nu'})}{(\epsilon_{{\bf k},\nu}-\epsilon_{{\bf k},\nu'})^2}\nonumber\\\label{eq:inter2}\end{align}
where $\nu=s,\eta$ for conciseness. We also set ${\bm\sigma}_\varsigma={\bm\sigma}(1+\varsigma\hat{\tau}_z)/2$, which defines the spin density operator on the spin-sublattice A ($\varsigma$=+1) and B ($\varsigma$=-1). Since Eqs. (\ref{eq:intra}) and (\ref{eq:inter2}) involve angular averaging over $\varphi_k$, it is convenient to evaluate the spin density in the limit $\theta\ll1$ (i.e. $\Lbhat\approx z$). In this case, the energy dispersion becomes isotropic $\epsilon_{s,\eta}=\eta(\sqrt{\gamma_k^2+J_{\rm sd}^2}+s\alpha k)$. \par

By taking the small $\alpha$ limit and replacing the discrete summation in Eqs. (\ref{eq:intra}) and (\ref{eq:inter2}) with continuous integration ($\sum_{\bf k}\rightarrow V\int d^2{\bf k}/4\pi^2$), one obtains in the linear order in $\alpha$ 
\begin{align}
{\bf S}^{\rm Intra}=&\frac{m\alpha}{8\pi\hbar^2\Gamma}\left(1+2\frac{J_{\rm sd}^2}{\epsilon_{\rm F}^2}\left[2-\frac{\epsilon_0}{\sqrt{\epsilon_{\rm F}^2-J_{\rm sd}^2}}\right]\right)({\bf z}\times e{\bf E}),\nonumber\\\label{eq: Sintaraf}\\
{\bf S}^{\rm Inter}=&-\varsigma\frac{m\alpha J_{\rm sd}}{4\pi \hbar^2\epsilon_{\rm F}^2}\left(1-\frac{\epsilon_0}{\sqrt{\epsilon_{\rm F}^2-J_{\rm sd}^2}}\right)(\Lbhat\times({\bf z}\times{\bf E})).\label{eq: Sinteraf}
\end{align}
where we defined $\epsilon_0=ta_l^2k_0^2$. These formulae hold for $\epsilon_{\rm F},J_{\rm sd}\gg\alpha k_F\gg\Gamma$ ($k_F$ is the Fermi wave vector). As predicted from symmetry considerations in the previous section, the intraband contribution produces an effective field along the vector ${\bf z}\times{\bf E}$, i.e. independent of the magnetic moments direction, while the (intrinsic) interband contribution results in a staggered effective field along the vector $\varsigma\Lbhat\times ({\bf z}\times{\bf E})$ that depends on the direction of magnetic moments and has opposite sign on the two spin-sublattices. These results are the AFM counterparts to the formulae obtained in the case of a FM 2D Rashba model \cite{Li15} and demonstrate that the torque enabling efficient electrical manipulation of the AFM order arises in this model from the odd interband contribution to the CISP which has a finite value in the $\Gamma \rightarrow 0$ limit, i.e., it is intrinsic in nature.

\subsection{Numerical calculations}

\begin{figure}[h]
  \includegraphics[width=8.6cm]{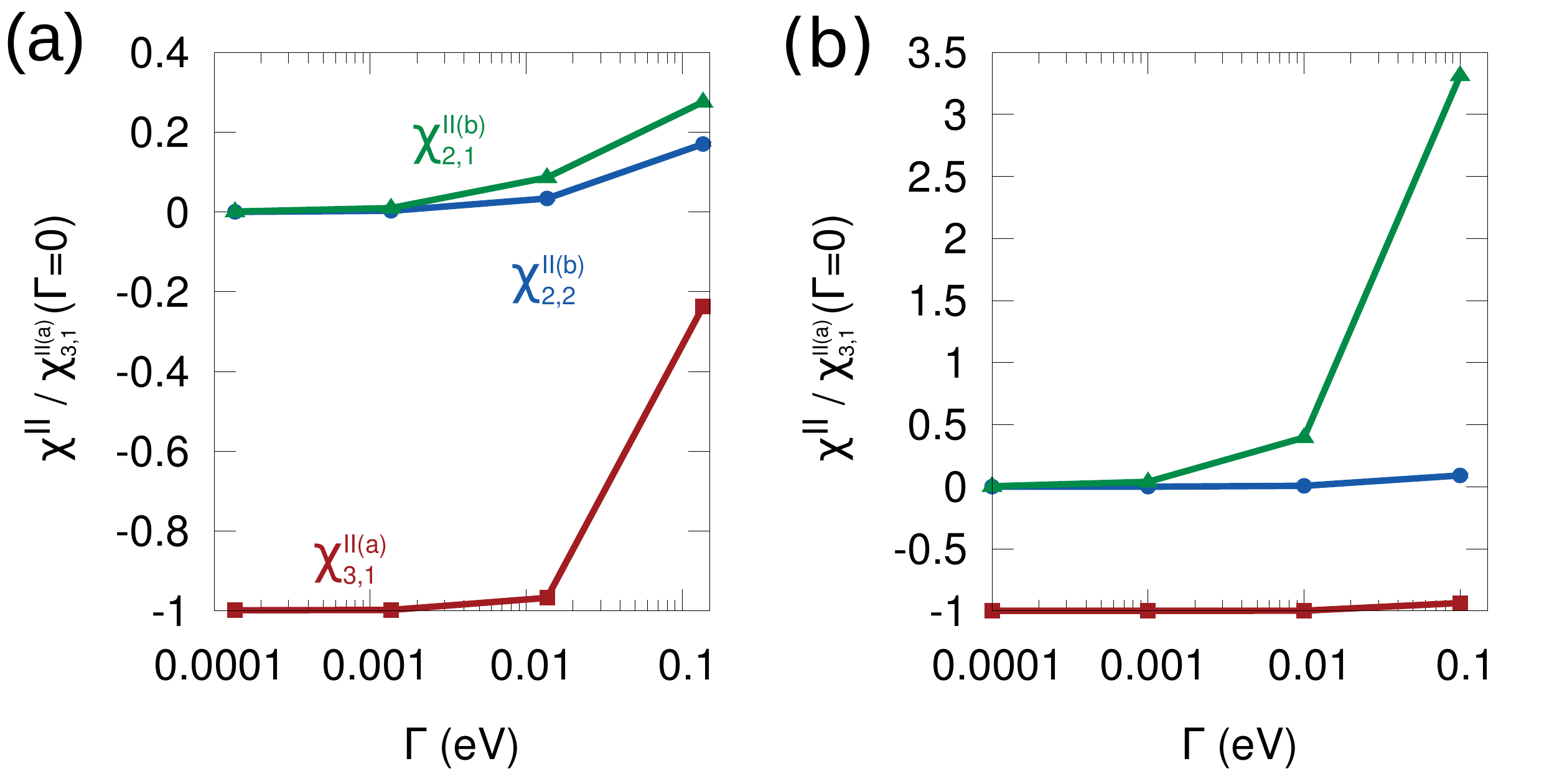}
  \caption{(Color online) $\Gamma$-dependence of terms $\chi^{II(a)}$, $\chi^{II(b)}$ in the two models for magnetic moments oriented along the [110] direction. Components that are not shown are zero or related to those shown by symmetry. The results are scaled to make the comparison between the two tight-binding models easier. (a) $\text{Mn}_2\text{Au}$, (b) the 2D model.}
  \label{fig:gamma}
\end{figure}

In this section we show results of numerical calculations of the CISP in the two models described in Section \ref{sec:models}. The intraband term $\chi_a^I$ for $\text{Mn}_2\text{Au}$ and the interband term  $\chi_a^{II(a)}$ for the 2D model were already presented in Ref. \cite{Zelezny2014}. Here we calculate also the interband term for $\text{Mn}_2\text{Au}$ and the intraband term for the 2D model. We are primarily interested in the small $\Gamma$ limit. For zero $\Gamma$ the interband term $\chi_a^{II(b)}$ vanishes. This is illustrated in Fig. \ref{fig:gamma}. Fig. \ref{fig:gamma}(a) shows the terms $\chi_a^{II(a)}$ and $\chi_a^{II(b)}$ as a function of $\Gamma$ for $\text{Mn}_2\text{Au}$ for magnetic moments oriented along the  [110] direction. Fig. \ref{fig:gamma}(b) shows the same calculation for the 2D model. In both cases as $\Gamma$ goes to zero, the term $\chi_a^{II(b)}$ goes to zero, while the term $\chi_a^{II(a)}$ becomes constant. In the following we choose $\Gamma$ so that the term $\chi_a^{II(b)}$ is small and the term $\chi_a^{II(a)}$ is close to its zero $\Gamma$ limit. We use $\Gamma\approx0.0013\ \text{eV}$ in $\text{Mn}_2\text{Au}$ and $\Gamma = 0.01\ \text{eV}$ in the 2D model. 

As discussed in the symmetry analysis in Sec. \ref{sec:symmetry}, the part of  CISP that is even under timer reversal is in the 2D model the same on the two AFM spin-sublattices, while the odd part of the CISP is opposite. Conversely in $\text{Mn}_2\text{Au}$, the even CISP is opposite and the odd CISP is the same. This is a key result since it shows that in both models the effective current-induced field has a staggered component and can therefore switch the AFM moments efficiently. In the following, we focus on the dependence of the CISP on the direction of magnetic moments. Since in our model systems the CISP is always either exactly the same or opposite on the two sublattices, we show results for one sublattice only.

We normalize all CISPs by current density calculated using the linear response theory formula analogous to \eqref{Boltzmann}. Since both the intraband term and the conductivity scale as $1/\Gamma$, the normalized intraband term is independent of $\Gamma$. For small $\Gamma$, the normalized term $\chi_a^{II(a)}$ scales as $\Gamma$. We also normalize the CISP by the ground-state spin-polarization (on each sublattice). When this quantity is multiplied by $J_{\text{sd}}/\mu_B$ we get directly the effective field.

\begin{figure}[h]
  \includegraphics[width=8.6cm]{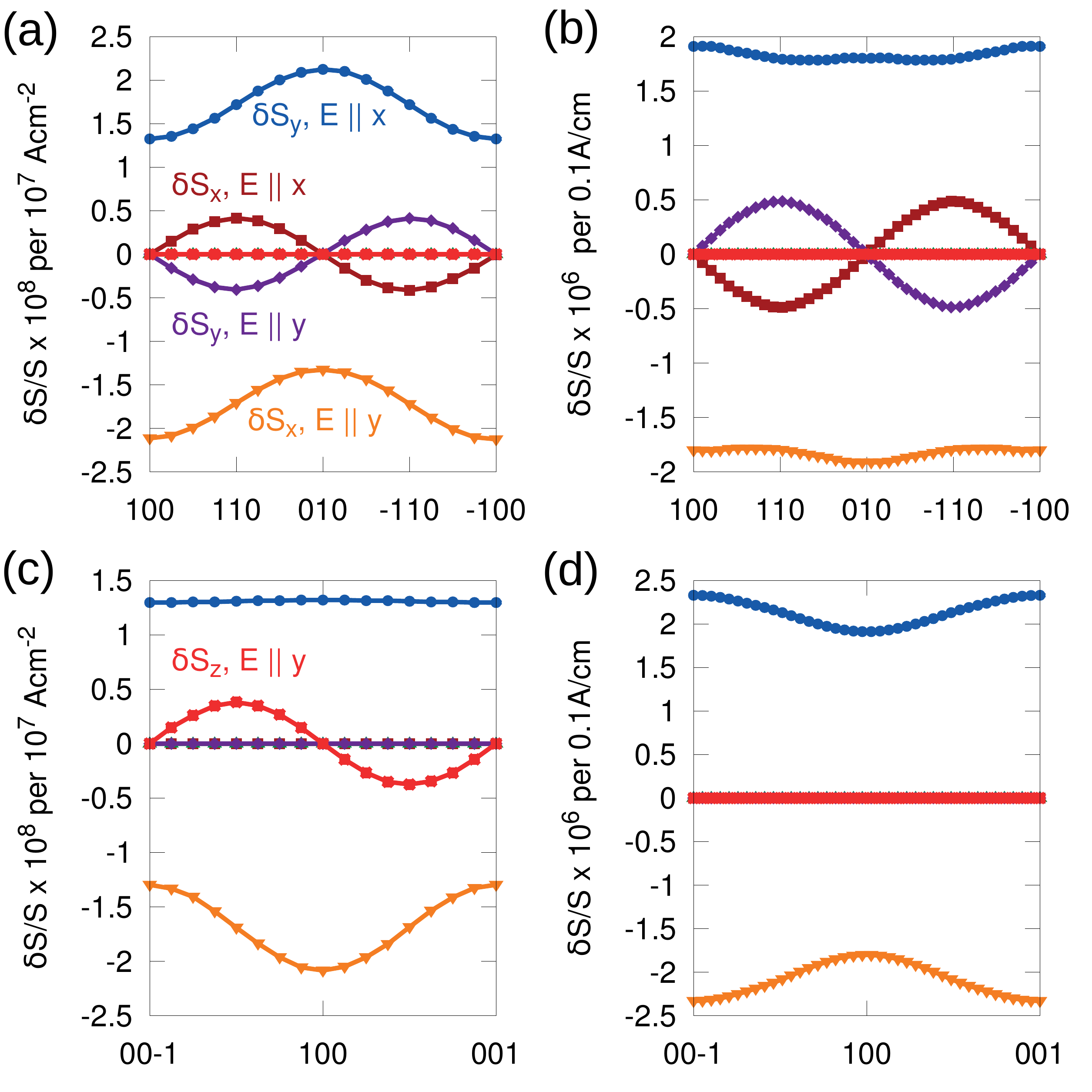}
  \caption{(Color online) Calculations of the intraband term $\chi_a^{I}$. Only results for one sublattice are shown. Plots show the intraband CISP normalized by the equilibrium spin-polarization per $10^7\ \text{Acm}^{-2}$ current density. (a) $\text{Mn}_2\text{Au}$ and in-plane rotation of magnetic moments \cite{Zelezny2014}, (b) 2D model, in-plane rotation, (c) $\text{Mn}_2\text{Au}$, out-of-plane rotation \cite{Zelezny2014}, (d) 2D model, out-of-plane rotation.}
  \label{fig:intra} 
\end{figure}

The results for the intraband term for one sublattice are shown in Fig. \ref{fig:intra}. For comparison we present the results for $\text{Mn}_2\text{Au}$ and the 2D model side by side. Figs.~\ref{fig:intra}(a),(c) show results for $\text{Mn}_2\text{Au}$, while Figs. \ref{fig:intra}(b),(d) show the 2D model. In Figs. \ref{fig:intra}(a),(b) the magnetic moments were rotated from the [100] direction to the [-100] direction through the [010] direction (the moments lie in-plane). In Figs. \ref{fig:intra}(c),(d) the magnetic moments were rotated from [00-1] direction to the [001] direction through the [100] direction (the moments lie out-of-plane). Only results for current along $x$ and $y$ directions are shown. For $\text{Mn}_2\text{Au}$, there can also be current along the $z$ direction, but we found that the CISPs for such a current are at least 2 orders of magnitude smaller than for the in-plane current. Note that the CISP for current in the $z$ direction is in general allowed by symmetry and only vanishes at certain high symmetry directions of magnetic moments. On one sublattice, the two models give qualitatively similar results. In both cases the CISP is not strongly dependent on the direction of magnetic moments and the dominant component is always in-plane and perpendicular to the current. The CISP and thus also the effective current-induced field are approximately aligned along the vector $\mathbf{z} \times \eb $. 

\begin{figure}[h]
  \includegraphics[width=8.6cm]{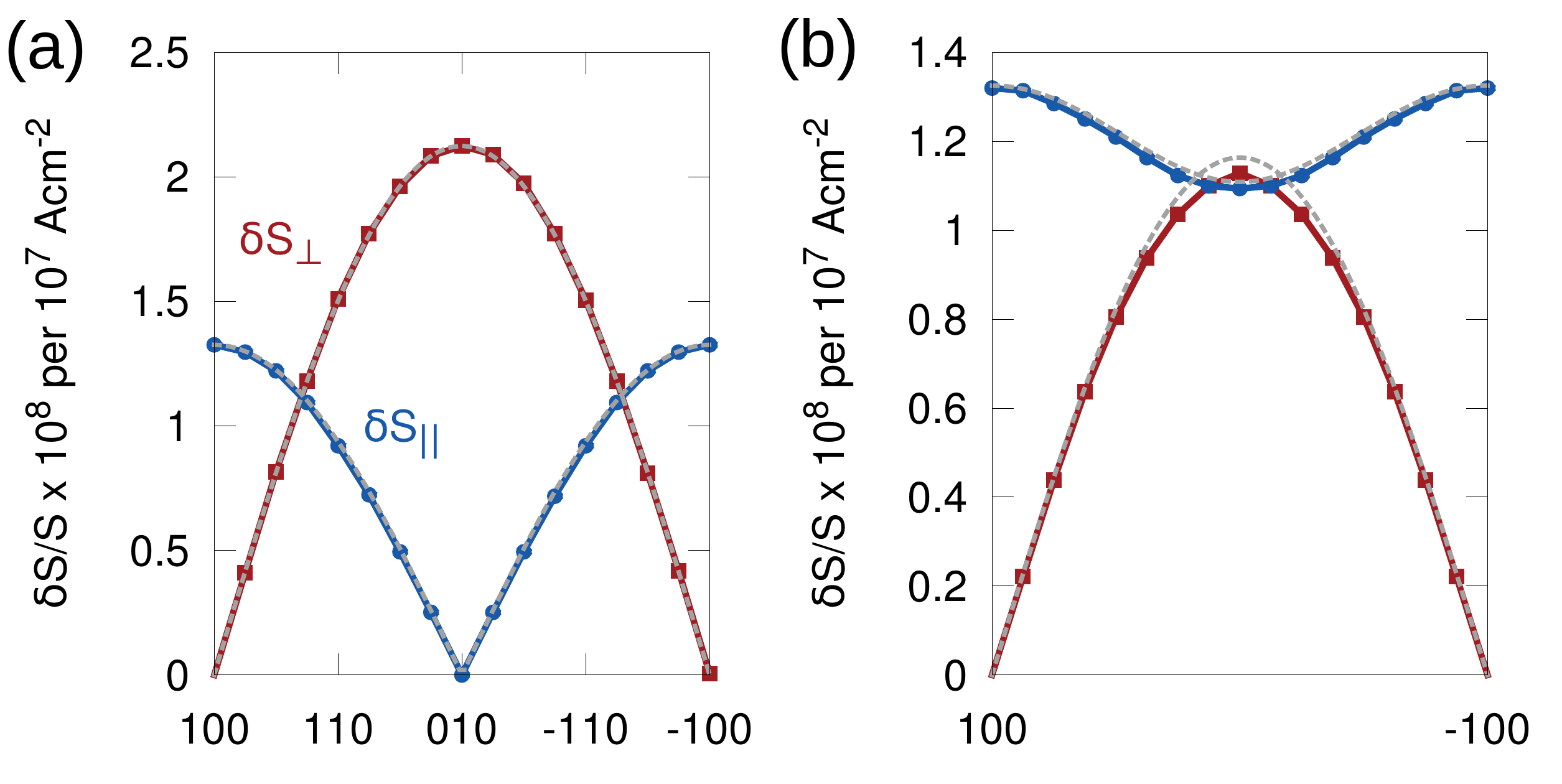}
  \caption{(Color online) Magnitude of perpendicular and longitudinal parts of the intraband CISP in $\text{Mn}_2\text{Au}$. Grey lines show perpendicular and longitudinal parts of the expression $\mathbf{z} \times \mathbf{E}$. (a) In-plane rotation of magnetic moments. (b) Non-symmetrical rotation: magnetic moments rotate along a path given by $\Lbhat = (\cos(\varphi),\sqrt{0.3}\sin(\varphi),\sqrt{0.7}\sin(\varphi))$}
  \label{fig:perpe} 
\end{figure}

Since the torque is a cross product of the effective field and the magnetic moment, only the component of the effective field which is perpendicular to the magnetic moment is relevant. In our models the effective field is proportional to the CISP so the same holds for the CISP. When the perpendicular component of the intraband CISP for $\text{Mn}_2\text{Au}$ is plotted, a peculiar feature is discovered. While the total CISP differs from the expression $\mathbf{z} \times \eb $ significantly, the perpendicular part is very close to the  perpendicular part of $\mathbf{z} \times \eb $. This is already manifested in Fig. \ref{fig:intra}(c), where the longitudinal component of the CISP is zero due to symmetry for current along the $x$ direction. To illustrate this feature we plot the magnitude of the perpendicular part of the intraband CISP for the in-plane rotation of the moments in  Fig. \ref{fig:perpe}(a). In gray, the perpendicular component of the expression $\mathbf{z} \times \eb $ is plotted. All  directions of magnetic moments discussed so far, lied in high symmetry planes. To confirm that this feature is not due to some particular symmetry, but rather a general feature of the model, we also rotated the moments along a non-symmetrical path. As shown in Fig. \ref{fig:perpe}(b) this rotation shows the same behavior.

Interestingly, the same holds for the longitudinal part of the CISP, i.e., the longitudinal part of the CISP is very close to the longitudinal part of $\mathbf{z} \times \eb $, which is also shown in Figs. \ref{fig:perpe}(a),(b). The proportionality constants are, however, different in the two cases which is why the total intraband CISP vector deviates from $\mathbf{z} \times \eb $. Since only the perpendicular part is relevant for the torque, the torque will be closely approximated by $\hat{\mathbf{M}}_a \times (\mathbf{z} \times \mathbf{E})$. This behavior only occurs in $\text{Mn}_2\text{Au}$. In the 2D model, the perpendicular component of CISP is not significantly closer to $\mathbf{z} \times \eb $ than the total CISP.

\begin{figure}[h]
  \includegraphics[width=8.6cm]{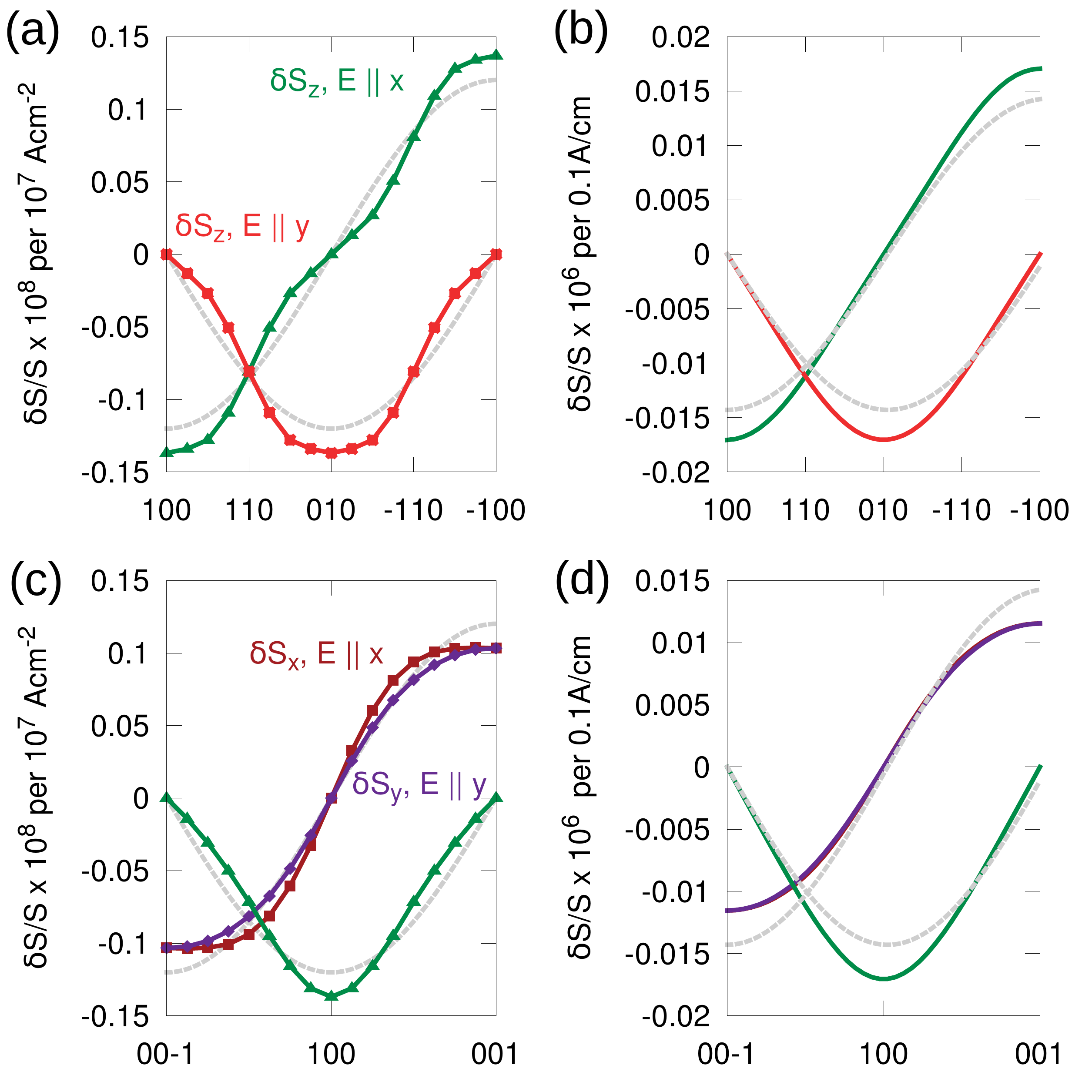}
  \caption{(Color online) Calculations of the interband term $\chi_a^{II}(a)$. Only results for one sublattice are shown. Plots show the interband CISP normalized by equilibrium spin-polarization per $10^7\ \text{Acm}^{-2}$ current density. Grey lines show a fit to the expression $\Lbhat \times (\mathbf{z} \times \mathbf{E})$. (a) $\text{Mn}_2\text{Au}$ and in-plane rotation of magnetic moments, (b) 2D model, in-plane rotation \cite{Zelezny2014}, (c) $\text{Mn}_2\text{Au}$, out-of-plane rotation, (d) 2D model, out-of-plane rotation \cite{Zelezny2014}.}
  \label{fig:inter}
\end{figure}

In Fig. \ref{fig:inter}, we show results for the interband term $\chi_a^{II(a)}$ for one sublattice, organized similarly to the intraband results in Fig. \ref{fig:intra}. The $\text{Mn}_2\text{Au}$ results are in Figs. \ref{fig:inter}(a),(c) and the 2D model results in Figs. \ref{fig:inter}(b),(d). In Figs. \ref{fig:inter}(a),(b) the CISP is plotted as a function of magnetic moments rotating in-plane, while in Figs. \ref{fig:inter}(c),(d), the magnetic moments are rotated out-of-plane. Again, the two models are qualitatively similar. In this case, however, the CISP depends strongly on the direction of magnetic moments. We only plot the non-negligible components of the CISP. In particular, the CISP for the current along the $z$ direction in $\text{Mn}_2\text{Au}$ is again very small. In both models the CISP can be closely approximated by the lowest order term  given by Eq. \eqref{eq:exp_odd}. As shown by grey lines, the main contribution is of the form $\Lbhat \times (\mathbf{z} \times \mathbf{E})$, which corresponds to the tensor $X_1$. A deviation from this form is mainly due to the presence of the tensor $X_2$. In $\text{Mn}_2\text{Au}$ also higher order terms are present, but are less important than the lowest order terms. In Mn2Au, the contribution from the tensor $X_3$ is also present, but we do not plot it since it is oriented approximately along the direction of magnetic moments and thus does not contribute to the torque.

%The contribution from the tensor $X_3$ is quite peculiar, as in $\text{Mn}_2\text{Au}$ it can exist even when there is no spin-orbit coupling. The CISP without spin-orbit coupling can, however, occur only in the direction of equilibrium magnetic moments. Therefore, without spin-orbit coupling it has to hold that $C_3=C_4$. We find that this contribution is indeed present and is comparable in magnitude to the other contributions that require spin-orbit coupling. Inclusion of spin-orbit coupling does not change this contribution much and it still approximately holds that $C_3 \approx C_4$. Because of this, the term $X_3$ has a negligible influence on the torque, and we thus don't plot it in Fig. \ref{fig:inter}. 

\begin{figure}[h]
  \includegraphics[width=8.6cm]{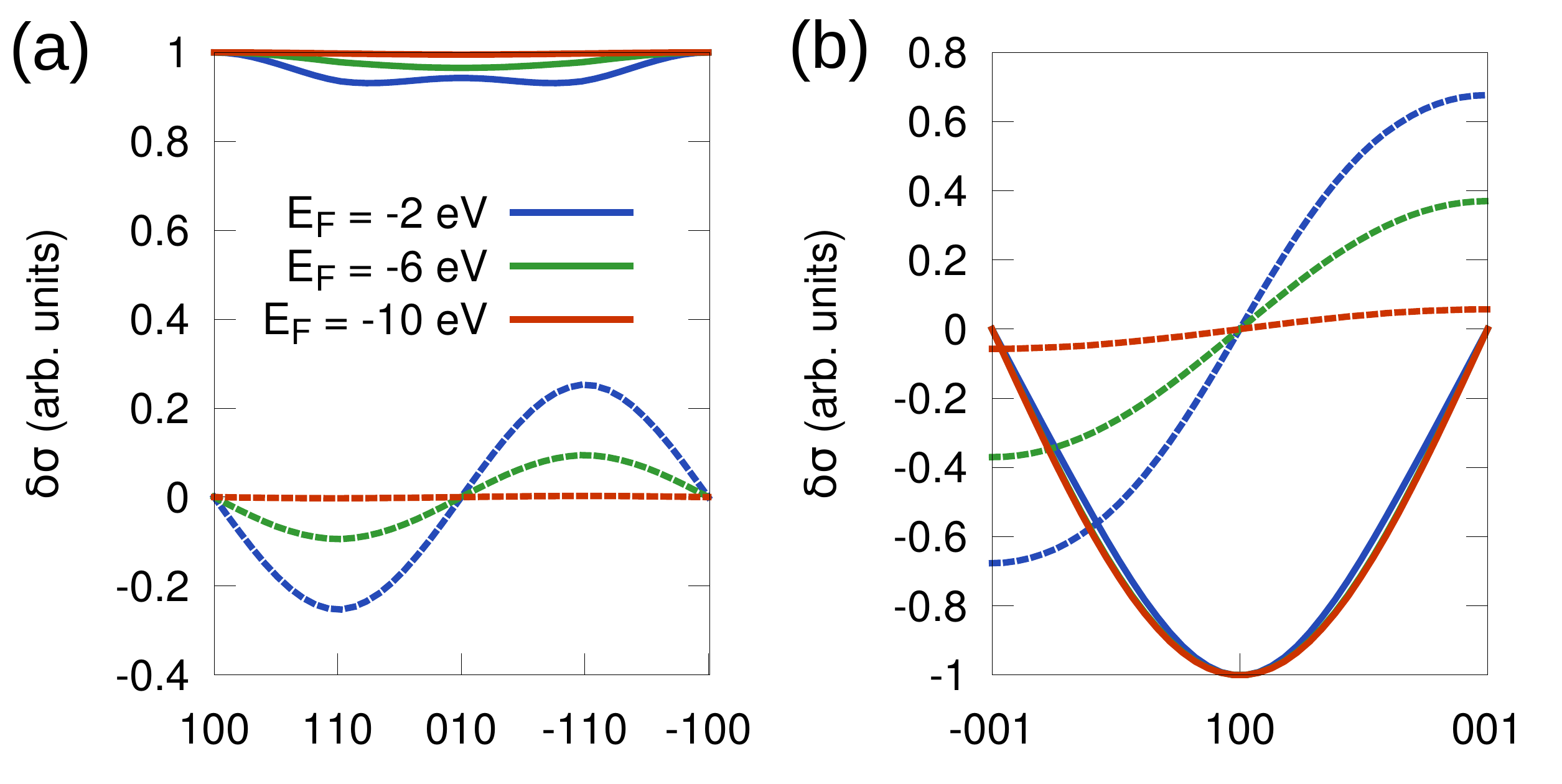}
  \caption{(Color online) The Dependence of the CISP in the AFM 2D Rashba model on the direction of magnetic moments for different Fermi levels. We scale results to highlight the change in shape. Only results for current along the $x$ direction are shown. $\Gamma=0.0001\ \text{eV}$ was used in this calculation.  (a) Intraband term, in-plane rotation, solid lines show the $y$ component, dashed lines show the $x$ component.(b) Interband term, out-of-plane rotation, solid lines show the $z$ component, dashed lines show the $x$ component. }
  \label{fig:fermi_level_dep}
\end{figure}

In Fig. \ref{fig:fermi_level_dep} we show how the CISP in the 2D model depends on the Fermi level. The dependence of the magnitude of the interband term on the Fermi level was already studied in Ref. \cite{Zelezny2014}. Here we focus instead on how the dependence of the CISP on the direction of magnetic moments changes when the Fermi level is varied. When the Fermi level approaches the bottom of the bands, the intraband term becomes independent of the direction of magnetic moments and can be described by the vector $\mathbf{z} \times \mathbf{E}$ very accurately. This is illustrated in Fig. \ref{fig:fermi_level_dep}(a). Results for the interband term $\chi_a^{II(a)}$ are shown in Fig. \ref{fig:fermi_level_dep}(b). For all Fermi level values, it can be described by Eq. \eqref{eq:exp_odd}, but the ratio $C_1/C_2$ depends strongly on the Fermi level. For the Fermi level close to the bandgap (see  Ref. \cite{Zelezny2014} for the bandstructure), $C_1$ is much larger than $C_2$. When the Fermi level approaches the bottom of bands, $C_2$ becomes much larger than $C_1$. The dependence of the  CISP on the direction of magnetic moments is then no longer of the form $\Lbhat \times (\mathbf{z} \times \mathbf{E})$. Instead, for $C_1<<C_2$, it can be described by $(\Lbhat\cdot \mathbf{E}_\perp)\mathbf{z}$. This is in agreement with the analytical calculations. Eq. \eqref{eq: Sinteraf} describes the contribution from the tensor $X_1$. When the Fermi level is at the bottom of bands the term given by Eq. \eqref{eq: Sinteraf} is zero. Eq. \eqref{eq: Sinteraf} does not capture the contribution from tensor $X_2$ since tensor $X_2$ is zero for $\Lbhat = \mathbf{z}$.

Finally, we compare results for our two models with AFM and FM order. Spin-orbit torques have been previously calculated in a FM 2D Rashba model analogous to our AFM 2D Rashba model \cite{Manchon2008,Manchon2009,Li15}. Those calculations used models with a parabolic band dispersion, which in our model corresponds to the Fermi level close to the bottom (top) of bands. As shown in Fig. \ref{fig:fermi_level_dep}(a), for our model the intraband CISP then becomes proportional to $\mathbf{z} \times \mathbf{E}$. This is a form that the FM has when $J_{\text{sd}}>>\alpha$ \cite{Li15}. We find that the AFM has this form regardless of $J_{\text{sd}}/\alpha$ (when the Fermi level is close to bottom of the bands). In all calculations discussed so far $\alpha << J_{\text{sd}}$. For such a case, the FM has the intraband term of the form $\mathbf{M}  \times [ ( \mathbf{z} \times \mathbf{E}) \times \mathbf{M}]$. This results in the same torque as the $\mathbf{z} \times \mathbf{E}$ term since  $\mathbf{M}  \times [ ( \mathbf{z} \times \mathbf{E}) \times \mathbf{M}]$ is precisely the component of $\mathbf{z} \times \mathbf{E}$ perpendicular to $\mathbf{M}$. The FM interband term $\chi_a^{II(a)}$ differs from the AFM case as well. For $\alpha << J_{\text{sd}}$ in the FM it has the form $\mathbf{M} \times (\mathbf{z} \times \mathbf{E})$, while for the AFM the dependence is $(\Lbhat\cdot \mathbf{E}_\perp)\mathbf{z}$, as shown in Fig. \ref{fig:fermi_level_dep}(b). This is a form the FM has when $J_{\text{sd}} >> \alpha$. 

The AFM model thus has many similarities with the FM model, however, the dependence on the parameters of the model is different. In particular, in the FM the results depend significantly on the ratio $J_{\text{sd}}/\alpha$, while in the AFM this ratio does not play a large role. This is because, in the FM, the spin-up and spin-down bands are split by both the Rashba spin-orbit coupling and by the exchange interaction. In the AFM on the other hand, only the Rashba spin-orbit coupling splits the spin-up and spin-down bands.

\begin{figure}[h]
  \includegraphics[width=8.6cm]{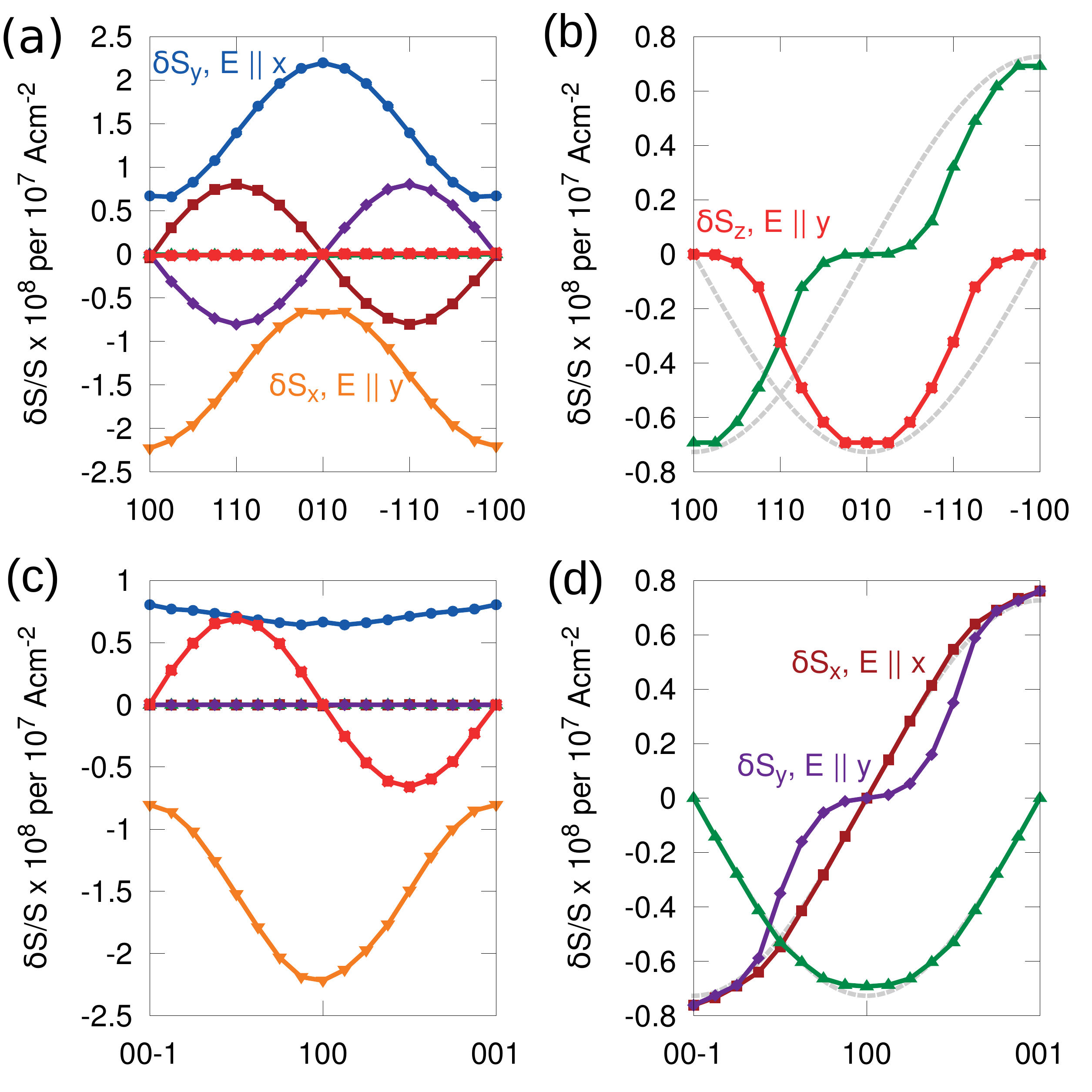}
  \caption{(Color online) Calculations of the CISP in FM $\text{Mn}_2\text{Au}$ for one inversion-partner sublattice. (a) Intraband term, in-plane rotation of moments. (b) Interband term $\chi_a^{II(a)}$, in-plane rotation of moments. (c) Intraband term, out-of-plane rotation of moments. (d) Interband term $\chi_a^{II(a)}$, out-of-plane rotation of moments.}
  \label{fig:FM}
\end{figure}

For comparison we also calculated the CISP in hypothetical FM $\text{Mn}_2\text{Au}$. The model differs from AFM $\text{Mn}_2\text{Au}$ only in the direction of the moments; all other parameters are the same. Both intraband and  interband CISPs in FM $\text{Mn}_2\text{Au}$ have opposite sign on the two inversion-partner lattice sites occupied by Mn, as expected from symmetry considerations and confirmed in our microscopic calculations. The intraband CISP in the FM is very close to the AFM case both in terms of the magnitude and the dependence on the direction of magnetic moments, as shown in Figs. \ref{fig:FM}(a),(c). The interband term is shown in Figs. \ref{fig:FM}(b),(d). It has a similar dependence on the direction of magnetic moments, however, is an order of magnitude larger than in the AFM $\text{Mn}_2\text{Au}$.  

\section{Discussion}
\label{sec:discussion}

The AFM 2D Rashba and 3D $\text{Mn}_2\text{Au}$ models differ in one key aspect. In the 2D model, the odd CISP is staggered and the even CISP is uniform, while in $\text{Mn}_2\text{Au}$ the even CISP is staggered and the odd CISP is uniform. This is so because in the 2D model the AFM spin-sublattices are connected by translation, while in $\text{Mn}_2\text{Au}$ they are connected by inversion. However, as shown in Figs. \ref{fig:intra} and \ref{fig:inter}, when we look at one sublattice only, the CISP in the two models has a similar dependence  on the direction of magnetic moments and the direction of the current.  This may seem surprising since the electronic structures of the two models are very different, including the way spin-orbit coupling enters the band-structure calculations. In the 2D model the spin-orbit coupling has the Rashba form, which effectively describes a structural inversion assymetry at, e.g., a surface/interface. It is 2D and has a unique vector associated with the model, which is normal to the surface. In $\text{Mn}_2\text{Au}$ on the other hand, the spin-orbit coupling that enters the microscopic tight-binding Hamiltonian is atomic and spherically symmetrical.

The reason for the similarity is the same local symmetry in the two models.  As discussed in Section \ref{sec:symmetry}, it is the local symmetry that determines the symmetry of the CISP on a sublattice. The local point group is the same in both models and in both models the results can be quite accurately described by lowest orders in expansion \eqref{eq:chi_expansion}. Since the local symmetry is the same, the expansions are also the same in the two models.

\begin{figure}[h]
  \includegraphics[width=8.6cm]{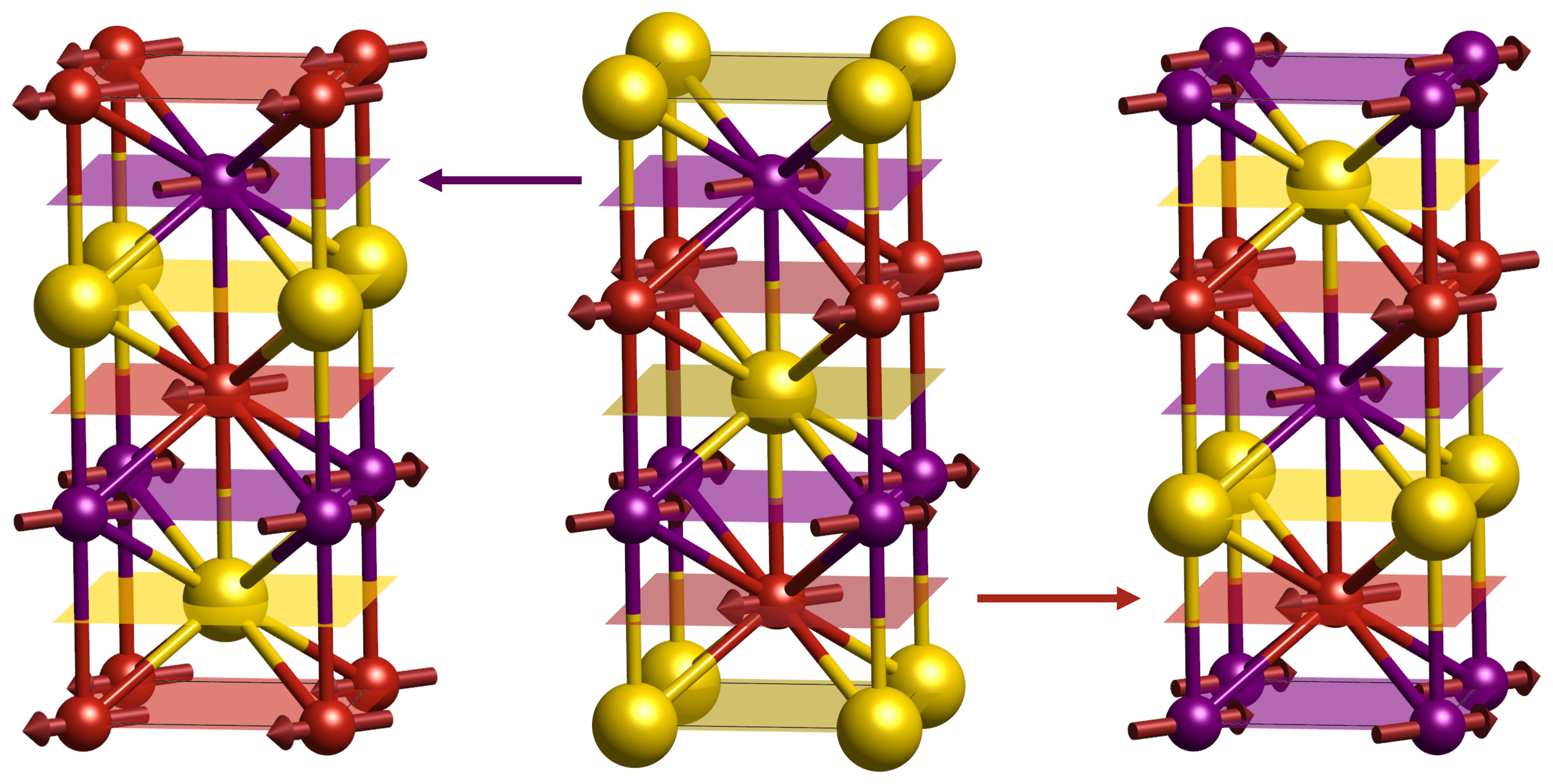}
  \caption{(Color online) Local inversion symmetry breaking in $\text{Mn}_2\text{Au}$. The middle picture shows the crystal structure of $\text{Mn}_2\text{Au}$ with highlighted atomic layers. Left and right pictures show inversion around Mn A and Mn B atoms (as defined in Fig. \ref{fig:structures}(b)) respectively. }
  \label{fig:Mn2Au_inv}
\end{figure}

The local inversion symmetry breaking in $\text{Mn}_2\text{Au}$ is illustrated in Fig. \ref{fig:Mn2Au_inv}. $\text{Mn}_2\text{Au}$ is a layered crystal; under inversion around one of the Mn atoms, the layers remain the same, but the order of the layers changes. Because of this, each sublattice has the inversion symmetry locally broken and the inversion symmetry breaking is along the $z$-axis. The inversion symmetry breaking thus resembles that of the 2D Rashba system.

%In AFM $\text{Mn}_2\text{Au}$, apart from the local inversion symmetry breaking, which is present even in the absence of magnetic order, there is also a global inversion symmetry breaking caused by the magnetic order. In FM $\text{Mn}_2\text{Au}$, only the local inversion symmetry breaking is present and yet the CISP in FM $\text{Mn}_2\text{Au}$ is very similar to that of AFM $\text{Mn}_2\text{Au}$. This holds especially for the even CISP, but the odd CISP is quite similar as well. This, together with the similarity to the 2D model shows that the torque in $\text{Mn}_2\text{Au}$ originates primarily from the local inversion symmetry breaking. The FM model also illustrates that the spin-orbit torque can occur even in material with global inversion symmetry. Such a torque will have very little effect on the FM though due to the exchange interaction.

Despite the above similarity in the local CISP symmetries of the two models, the dynamics of magnetic moments will be different. This is because only the staggered component of the CISP can generate an efficient torque on an AFM. In $\text{Mn}_2\text{Au}$, the staggered component of the CISP corresponds to the intraband term, which results in a field-like torque. The effect of such a torque is comparable to the effect of an external magnetic field in a FM. In particular, the magnitude of the staggered effective field necessary to switch the AFM moments will be determined by the magnetic anisotropy energy barrier just like in the case of a FM and a uniform external magnetic field. In the 2D model, the staggered component of the CISP depends strongly on the magnetic moments direction and is thus not field-like. When the Fermi level is such that the CISP has the $\Lbhat \times ( \mathbf{z} \times \mathbf{E})$ form, the corresponding torque acting on the AFM can be called, in analogy with FMs,  (anti)damping-like. The critical value of the switching effective field will then depend both on the anisotropy barrier and on the damping factor \cite{Gomonay2012}.

These results demonstrate the importance of symmetry for understanding the spin-orbit torque. Symmetry determines which component of the effective field is staggered and thus also which component is efficient for manipulating the AFM order. Symmetry also governs the dependence of the effective field on the direction of the current and the magnetic moments. This is especially so because we find in our two models that the effective field can be very well described by the lowest order terms in the expansion \eqref{eq:chi_expansion}. Although this conclusion does not have to be generally valid, it is consistent with previous studies on different systems \cite{Fang2011,Kurebayashi2014,Wadley2016,ciccarelli2015}.

We used the tight-binding models of the 3D $\text{Mn}_2\text{Au}$  and 2D Rashba AFMs to illustrate the symmetries of CISPs and the corresponding AFM spin-orbit torques. In remaining paragraphs  we discuss the strength of the spin-orbit torque in the $\text{Mn}_2\text{Au}$ crystal. In  Ref. \cite{Zelezny2014}, the magnitude of the effective field driving the spin-orbit torque in $\text{Mn}_2\text{Au}$ was estimated from the tight-binding value of the CISP by estimating the exchange coupling strength between carrier and local moment spins to $J_\text{sd}=1\,\text{eV}$. The intraband field was found to be 0.22 mT per $10^7\ \text{Acm}^{-2}$ for magnetic moments lying along the [100] direction. For comparison, we performed ab initio calculations based on the electronic structure obtained from the DFT. The method is described in detail in Ref. \cite{freimuth2014}. Here the spin-orbit torque is calculated directly using the exchange potential from the DFT. From the torque the effective magnetic field is then obtained using Eq. \eqref{eq:torque}: $\mathbf{B}_a = (\mathbf{T}_a \times \hat{\mathbf{M}}_a)/ M_a$. This way we only obtain the component of the effective field perpendicular to the magnetic moments. For magnetic moments along the [100] direction the longitudinal component of the effective field is zero. We found that the effective current induced field in the ab initio spin-orbit torque calculation has a magnitude  1.98 mT per  $10^7 \text{Acm}^{-2}$ \cite{Wadley2016}. 

The DFT value is by a factor of 8 larger than the tight-binding value. To find out where the discrepancy originates from, we also calculated the CISP directly using the DFT method. Then using Eq. \eqref{eq:eff_field} and the above DFT torque calculation we obtain a value of the effective exchange constant $J_\text{sd}=1.2$~eV. This is similar to the estimated value of $J_\text{sd}$ used in the tight-binding calculation of the effective current induced field. The difference between tight-binding and DFT calculations is therefore primarily in the CISPs, which differ by a factor of $\sim 6$. The remaining discrepancy between the tight-binding and DFT effective fields is due to different magnetic moments in the two approaches. These differ because the tight-binding Hamiltonian was fitted to a LDA+U DFT calculation, while for the torque calculation a DFT calculation without U was used. Including the Hubbard U increases the moments by about 20\%.

The DFT magnitude of the effective staggered field in $\text{Mn}_2\text{Au}$ is comparable to that of the CuMnAs AFM where current induced switching has been recently observed in experiment \cite{Wadley2016}. Highly conductive $\text{Mn}_2\text{Au}$  is therefore a potentially  favorable material for exploring and exploiting current induced spin-orbit torques is AFMs.

\section{Conclusion}
\label{sec:conclusions}

We have presented a symmetry analysis of spin-orbit torques in AFMs and FMs and discussed in detail results obtained in two complementary model systems with locally and globally broken inversion symmetry, respectively. We have pointed out that the existence and form of the spin-orbit torque on the given spin-sublattice is determined by the sublattice local symmetry. We have also shown that in AFMs, symmetry operations that connect the two spin-sublattices determine the relation between the spin-orbit torques on the two sublattices. Our two models illustrate two main cases with the sublattices connected either by a translation or by an inversion. Consequently, in the AFM 2D Rashba model representing the former case, the efficient component of the torque has an antidamping character, while in AFM $\text{Mn}_2\text{Au}$ representing the latter case, the efficient spin-orbit torque has a field-like character.

\begin{acknowledgments}
We acknowledge support from the European Union (EU) European
Research Council Advanced (grant 268066); the Ministry of
Education of the Czech Republic (grant LM2011026); the Grant
Agency of the Czech Republic (grant 14-37427); Deutsche Forschungsgemeinschaft SPP 1538; supercomputing resources at J\"ulich Supercomputing Centre and RWTH Aachen
University;  DFG Transregional Collaborative
Research Center (SFB/TRR) 173 “Spin+X Spin in its collective environment” and the Alexander von Humboldt Foundation. Access to computing and storage facilities owned by parties and projects contributing to the National Grid Infrastructure MetaCentrum, provided under the programme "Projects of Large Research, Development, and Innovations Infrastructures" (CESNET LM2015042), is greatly appreciated. A.M. acknowledges financial support from the King Abdullah University of Science and Technology (KAUST) Office of sponsored Research (OSR) under Award OSR-2015-CRG4-2626 as well as fruitful discussions with H.B.M. Saidaoui. 
\end{acknowledgments}

\appendix
\section{Symmetry of spin-orbit torque}\label{appendix:symmetry}
We give here an overview of the symmetry of spin-orbit torque. Symmetry of transport coefficients in magnetic systems has been studied before, primarily in the context of electrical and heat conductivity \cite{Shtrikman1965,Kleiner1966,Seemann2015,Grimmer1993}, but also for the spin-orbit torque \cite{Wimmer2016}. Here we expand the analysis to account for sublattice projections. We use the microscopic Eqs. \eqref{Boltzmann},\eqref{eq:Kubo1},\eqref{eq:Kubo2} as a starting point. Our approach is similar to that of \cite{Kleiner1966} where a more general version of Kubo formula was used.  The results do not depend on the exact form of the formulas: it is only important that the formulas describe linear response. Additionally, the results apply only assuming a single-electron (i.e., non-interacting) Hamiltonian. 

As discussed in \cite{Grimmer1993}, there has been a considerable controversy surrounding the symmetry of tensors describing transport phenomena. The difficulty lies in understanding the effect of the time-reversal symmetry operation \cite{BUTZAL1982518,Grimmer1993}. This is because transport phenomena are non-equilibrium processes that include dissipation. We define a time-reversal operator as $\mathcal{T}=i\sigma_yK$, where $K$ is the complex conjugation. This is how the time reversal operator is conventionally defined in quantum mechanics. Note that such defined time-reversal operator reverses direction of magnetic moments, but does not in general reverse direction of electrical currents (see the discussion in \cite{BUTZAL1982518}).

Let $R$ be a symmetry of the Hamiltonian, i.e.,
\begin{align}
 H=RHR^{-1}.
\end{align}
Symmetry operations that do not contain time-reversal are represented by a unitary $R$. Symmetry operations that contain time-reversal are represented by antiunitary $R$ since $K$ is an antiunitary operator. If $\psi_{n\mathbf{k}}$ is an eigenfunction of the Hamiltonian, then $R\psi_{n\mathbf{k}}$ is also an eigenfunction with the same eigenvalue. Since the result cannot depend on the choice of eigenfunctions, we can use the transformed eigenfunctions to evaluate the CISP. The transformed eigenfunction correspond to a different $\mathbf{k}$-point, but the sums will always run over all $\mathbf{k}$-points. The only part of the microscopic equations that depends on the eigenfunctions are the matrix elements, the rest depends only on the eigenvalues.  Transformation of the matrix elements depends on whether $R$ is a unitary operator or an antiunitary operator. For unitary $R$ and an observable $\hat{A}$
\begin{align}
  \Bra{R(\psi_{n\mathbf{k}})}\hat{A}\Ket{R(\psi_{m\mathbf{k}})} &=  \Bra{\psi_{n\mathbf{k}}}R^{-1}\hat{A}R\Ket{\psi_{m\mathbf{k}}},
\end{align}
while for antiunitary $R$
\begin{align}
  \Bra{R(\psi_{n\mathbf{k}})}\hat{A}\Ket{R(\psi_{m\mathbf{k}})} &=  \Bra{\psi_{n\mathbf{k}}}R^{-1}\hat{A}R\Ket{\psi_{m\mathbf{k}}}^*
\end{align}	
We represent the transformation of operators $\hat{\mathbf{S}}$ and $\hat{\mathbf{v}}$ by 3x3 matrices $D^s$, $D^v$
\begin{align}
 R^{-1}\hat{S}_{a',i} R &= D_{ik}^{s} \hat{S}_{a,k},\\
 R^{-1}\hat{v}_j R &= D_{jl}^{v} \hat{v}_l,
\end{align}
where $a'$ is the sublattice in which $a$ transforms under $R$. Note that the matrix $D^s$ does not depend on $a$. For unitary $R$ we find for the transformation of $\chi_a$
\begin{align}
 \chi_{a',ij} = D_{ik}^{s}D_{jl}^{v}\chi_{a,kl}.\label{eq:trans_unitary},
\end{align}

For antiunitary $R$, the various terms of \eqref{eq:Kubo} will transform differently depending on whether they contain a real or imaginary part of the matrix elements.  To group together the terms that transform in the same way, we separate the spin-polarization into parts even and odd under time-reversal. Since time-reversal switches the direction of all moments, this is equivalent to
\begin{align}
 \chi_a^{\text{even}}([\mathbf{M}]) = \big[\chi_a([\mathbf{M}]) + \chi_a([-\mathbf{M}])\big]/2,\\
 \chi_a^{\text{odd}}([\mathbf{M}]) = \big[\chi_a([\mathbf{M}]) - \chi_a([-\mathbf{M}])\big]/2,
\end{align}
where $[\mathbf{M}] = [\mathbf{M}_A,\mathbf{M}_B,\dots]$ denotes the directions of all magnetic moments in the unit cell. Since both $\hat{\mathbf{S}}$ and $\hat{\mathbf{v}}$ anticommute with time-reversal and since $\chi_a^{I}$, $\chi_{a}^{II(b)}$ contain the real part of the matrix elements, while $\chi_a^{II(a)}$ contains the imaginary part of the matrix elements
\begin{align}
 \chi_a^{\text{even}} &= \chi_a^{I}+\chi_{a}^{II(b)},\\
 \chi_a^{\text{odd}} &= \chi_{a}^{II(a)}.
\end{align}
We find for the transformation of each part under antiunitary $R$
\begin{align}
 \chi_{a',ij}^{\text{even}} &= D_{ik}^{s}D_{jl}^{v}\chi_{a,kl}^{\text{even}},\label{eq:trans_even}\\
 \chi_{a',ij}^{\text{odd}} &= -D_{ik}^{s}D_{jl}^{v}\chi_{a,kl}^{\text{odd}}.\label{eq:trans_odd}
\end{align}
We now show how to express the matrices $D^s, D^v$. Let $D$ be a $3\times3$ matrix that represents how a point transforms under $R$
\begin{align}
 \mathbf{x}_R = D\mathbf{x}+\mathbf{s}\label{eq:matrix_D}.
\end{align}
The shift $\mathbf{s}$ is due to translations. It is irrelevant for matrices $D^s, D^v$, but the translations cannot be ignored altogether because they influence which sublattice $a$ transforms to. Note that the time-reversal symmetry operation does not influence the matrix $D$ either since it only affects the magnetic moments. Thus $D$ represents the nonmagnetic point group.

For unitary $R$
\begin{align}
D^s = \det(D)D, && D^v = D,
\end{align}
and for antiunitary $R$
\begin{align}
D^s = -\det(D)D, && D^v = -D.
\end{align}
This is because $\hat{\mathbf{v}}$ is a polar vector, while $\hat{\mathbf{S}}$ is an axial vector and both change sign under time-reversal. Then Eq. \eqref{eq:trans_unitary} can be rewritten as
\begin{align}
 \chi_{a'} = \det(D)D\chi_{a}D^T \label{eq:trans_unitary_mat}
\end{align}
and analogously for antiunitary $R$ 
\begin{align}
 \chi_{a'}^{\text{even}} &= \det(D)D\chi_{a}^{\text{even}}D^{T},\label{eq:trans_even_mat}\\
 \chi_{a'}^{\text{odd}} &= -\det(D)D\chi_{a}^{\text{odd}}D^T.\label{eq:trans_odd_mat}
\end{align}

Eqs. \eqref{eq:trans_unitary_mat}, \eqref{eq:trans_even_mat}, \eqref{eq:trans_odd_mat} together determine the transformation properties of the tensor $\chi_a$.  They hold, however, only in a cartesian coordinate system. This is because the formulas \eqref{Boltzmann}, \eqref{eq:Kubo1}, \eqref{eq:Kubo2} hold only in a cartesian system. In any coordinate system, for example, the CISP corresponding to the term $\chi_a^I$ can be written as
\begin{align}
 \delta \mathbf{S}_a = -\frac{e\hbar}{2\Gamma}\sum_{\kb,n} 
 \Bra{\psi_{n\mathbf{k}}} &\hat{\mathbf{S}}_a \Ket{\psi_{n\mathbf{k}}}
 \Bra{\psi_{n\mathbf{k}}} \hat{\mathbf{v}}\cdot \mathbf{E} \Ket{\psi_{n\mathbf{k}}}\notag\\
 &\times \delta(\varepsilon_{\mathbf{k}n}-E_F)
\end{align}
The terms corresponding to $\chi_a^{II(a)},\chi_a^{II(b)}$ can be expressed analogously. In a non-cartesian coordinate system, tensor $\chi_a$ would not satisfy $\delta \mathbf{S}_a = \chi_a \mathbf{E}$ since in a non-cartesian coordinate system $\hat{\mathbf{v}}\cdot \mathbf{E} \neq \hat{v}_iE_i$.  While, it is natural to express the tensors $\chi$ in a cartesian coordinate systems, the symmetry operations are usually expressed in the conventional coordinate systems, which for the case of monoclinic, hexagonal and trigonal groups are not cartesian. For completeness we provide here a generalization to non-cartesian systems. This can be derived by using microscopic expression for $\chi_a$ valid in non-cartesian systems, but a simpler way is to transform the linear response tensor from a non-cartesian system to cartesian. 

Let $T$ be a coordinate transformation matrix, i.e. a matrix such that $x'=Tx$, where $x$ are coordinates in a cartesian system and $x'$ are coordinates in a different, in general non-cartesian, coordinate system. Then $ \chi_a = T^{-1}\chi'_aT $. We first consider a unitary symmetry operation $R$. Using the Eq. \eqref{eq:trans_unitary_mat}, which holds in the cartesian system
\begin{align}
 T^{-1}\chi'_{a'}T &= \det(D) D T^{-1}\chi'_a T D^T \\
 \chi'_{a'} &= \det(D) T D T^{-1}\chi'_a T D^T T^{-1}
\end{align}
In a cartesian system, $D$ has to be orthogonal, so $D^{T}=D^{-1}$. Since $D'=TDT^{-1}$ and $\det(D')=\det(D)$, we find
\begin{align}
 \chi'_{a'} &= \det(D') D' \chi'_a D'^{-1}\label{eq:trans_fa1}.
\end{align}
Analogously, we find for antiunitary $R$
\begin{align}
 \chi'^{\text{even}}_{a'} &= \det(D') D' \chi'^{\text{even}}_a D'^{-1},\\
 \chi'^{\text{odd}}_{a'} &= -\det(D') D' \chi'^{\text{odd}}_a D'^{-1}.
\end{align}
These formulas determine how $\chi$ transforms in any coordinate system. This result is quite general and holds for any linear response formula. One just has to replace the matrices $D^s$, $D^v$ by matrices that describe transformation of the corresponding operators. 

The results for cartesian coordinate systems are the same as in Ref. \cite{Kleiner1966}, except that we separate the tensor into the even and odd parts. Such separation is also commonly done for other tensors describing transport phenomena \cite{Shtrikman1965,Grimmer1993}. It is quite natural since the even and the odd parts have different properties. For example, they have a different dependence on disorder and cause very different magnetic dynamics.

To find out symmetry properties of the expansion terms in \eqref{eq:chi_expansion} we have to consider the symmetry operations of the nonmagnetic crystal, since these are symmetry operations that connect different magnetic configurations of a given crystal. If $R$ is such symmetry operation then $H([\mathbf{M}]_R)=RH([\mathbf{M}])R^{-1}$, where $[\mathbf{M}]_R$ denotes directions of all magnetic moments transformed by $R$. By using a completely similar procedure as for deriving Eq. \eqref{eq:trans_fa1}, we can show that 
\begin{align}
  \chi_{a'}([\mathbf{M}]_R) &= \det(D) D\chi_{a}([\mathbf{M}])D^{-1}\label{eq:trans_mag}.
\end{align}
Since the nonmagnetic symmetry operations do not contain time-reversal we do not have to separate $\chi_a$ into the even and odd parts. Considering that Eq. \eqref{eq:trans_mag} has to hold for each expansion term in \eqref{eq:chi_expansion} we find
\begin{align}
 \chi_{a,ij,mn\dots}^{(i)} = \det(D)^{i-1}D_{ik}D_{jl}^{-T} D_{mo}^{-T} D_{np}^{-T}\dots \chi_{a,kl,op\dots}^{(i)},\label{eq:trans_expansion}
\end{align}
where $D^{-T}$ denotes the inverse of a transpose of matrix $D$. We consider here only the symmetry operations that keep sublattice invariant. The symmetry operations that connect the two sublattices do not give any more information about the form of $\chi_a^{(i)}$. The form of the expansion \eqref{eq:chi_expansion} is thus given by the nonmagnetic local point group.

In FMs Eq. \eqref{eq:trans_expansion} applies also for the expansion of the net CISP, if the global point group is used instead. In AFMs, the net tensor $\chi$ transforms in general differently. This is because a symmetry operation that transforms one sublattice into the other can reverse the sign of the N\'eel vector and this is not taken into account in Eq. \eqref{eq:trans_expansion}. For example in $\text{Mn}_2\text{Au}$, inversion is a symmetry of the nonmagnetic crystal. As a consequence, the net CISP in the FM $\text{Mn}_2\text{Au}$ vanishes as correctly predicted by Eq. \eqref{eq:trans_expansion} applied for the net CISP. In AFM $\text{Mn}_2\text{Au}$, there is, however, a net CISP, yet Eq. \eqref{eq:trans_expansion} is the same as for the FM. It is straightforward to derive the analogue of Eq. \eqref{eq:trans_expansion} for net CISP in AFMs, however, in AFMs the net CISP is not a very useful quantity. In AFMs with more than two sublattices, spin-axis direction is not enough to describe the magnetic state of the AFM. Then the expansion \eqref{eq:chi_expansion} should be performed in more parameters than just $\hat{\mathbf{n}}$. However, if we assume that all the other parameters are fixed during dynamics then expansion \eqref{eq:chi_expansion} can still be used and Eq. \eqref{eq:trans_expansion} applies.

\section{The code for analyzing the symmetry}\label{appendix:code}
We provide a code that can analyze the symmetry of spin-orbit torque in a given crystal structure automatically \cite{symcode}. It can find the symmetry restricted form of a tensor $\chi_a$ and also of the expansion \eqref{eq:chi_expansion}. Here, we give a brief overview of the code.  The code is written in Python and uses a symbolic library Sympy \cite{sympy}. It uses as an input a list of symmetry operations for the given crystal structure, generated by the program Findsym \cite{findsym}. From these symmetry operations we first choose the ones which form the local point group, i.e., the symmetry operations that leave the selected sublattice invariant. For each such symmetry operation we then construct a system of linear equations \eqref{eq:trans_f1}, \eqref{eq:trans_f2} (resp. \eqref{eq:trans_expansion} for the expansions) that have to hold for components of the tensor. We solve this system of equation by transforming it to the reduced row echelon form. The code can also find relations between tensors $\chi_a$ projected on different sublattices and between tensors $\chi_a$ for different equivalent magnetic configurations.

\input{refs.bbl}
% Create the reference section using BibTeX:
%\bibliography{/home/kuba/SparkleShare/Work/jabref/all.bib,SOT.bib}

\end{document}

%% file: refs.bbl
%merlin.mbs apsrev4-1.bst 2010-07-25 4.21a (PWD, AO, DPC) hacked
%Control: key (0)
%Control: author (8) initials jnrlst
%Control: editor formatted (1) identically to author
%Control: production of article title (-1) disabled
%Control: page (0) single
%Control: year (1) truncated
%Control: production of eprint (0) enabled
%